\documentclass[11pt,reqno]{amsart}

\usepackage[utf8]{inputenc}

\usepackage{amsmath,amsthm,amssymb}
\usepackage{amssymb}

\usepackage{graphics}
\usepackage{hyperref}
\usepackage[usenames, dvipsnames]{xcolor}

\usepackage{mathtools}
\mathtoolsset{showonlyrefs}

\usepackage[square,sort,comma,numbers]{natbib}

\definecolor{darkblue}{rgb}{0.0,0.0,0.3}
\hypersetup{colorlinks,breaklinks,
  linkcolor=darkblue,urlcolor=darkblue,
anchorcolor=darkblue,citecolor=darkblue}

\usepackage{booktabs}
\usepackage{threeparttable}

\theoremstyle{plain}
\newtheorem{theorem}{Theorem}
\newtheorem*{theorem*}{Theorem}

\newtheorem*{proposition*}{Proposition}

\newtheorem*{corollary*}{Corollary}

\newtheorem*{conjecture*}{Conjecture}

\theoremstyle{definition}
\newtheorem{remark}[theorem]{Remark}

\DeclareMathOperator{\SL}{SL}

\title{Visualizing Modular Forms}
\author{David Lowry-Duda}
\thanks{This work was supported by the Simons Collaboration in Arithmetic
Geometry, Number Theory, and Computation via the Simons Foundation grant 546235.}
\thanks{The author benefitted from many conversations at the fall 2019 ICERM
program on Illustrating Mathematics (where this project was born) and at the
special number theory day organized at Bowdoin College by Naomi Tanabe.}

\begin{document}

\maketitle

\begin{abstract}

We examine several currently used techniques for visualizing complex-valued
functions applied to modular forms. We plot several examples and study the
benefits and limitations of each technique. We then introduce a method of
visualization that can take advantage of colormaps in Python's
\texttt{matplotlib} library, describe an implementation, and give more
examples. Much of this discussion applies to general visualizations of
complex-valued functions in the plane.

\end{abstract}

\section{Introduction}

\subsection{Motivation}

Graphs of real-valued functions are ubiquitous and commonly serve as a source of
mathematical insight. But graphs of complex functions are simultaneously less
common and more challenging to make. The reason is that the graph
of a complex function is naturally a surface in four dimensions, and there are
not many intuitive embeddings available to capture this surface within a 2d
plot.

In this article, we examine different methods for visualizing plots of
modular forms on congruent subgroups of $\SL(2, \mathbb{Z})$. These forms are
highly symmetric functions and we should expect their plots to capture many
distinctive, highly symmetric features.

In addition, we wish to take advantage of the broader capabilities that exist in
the Python/SageMath data visualization ecosystem. There are a vast number of
color choices and colormaps implemented in terms of Python's \texttt{matplotlib}
library~\cite{matplotlib}. While many of these are purely aesthetic, others are
chosen to allow the most accurate perception by as many viewers as possible.
By using \texttt{matplotlib} colormaps, it is possible to
choose colors to support different types of color perception, including for
people who see color differently. We describe this further
in~\S\ref{sec:color}.

Although our emphasis is on plots of modular forms, many of the techniques
described apply more generally to graphing complex-valued functions. We hope
this paper will be of help to any reader wanting to visualize complex
functions.

\subsection{Broad Overview of Complex Function Plotting}

Over the last 20 years, different approaches towards representing
graphs of complex functions have emerged. The most commonly used
approach is one first introduced by Frank Farris in a
review~\cite{farris1998visual} of Needham's \emph{Visual Complex
Analysis}.

The idea is to represent the output of a complex function though color.
Frequently, one uses hue to represent argument and lightness to
represent magnitude. Or more generally, one can associate a color to each
point in the complex plane and then color the domain of a complex function $f$ by
the color of $f(z)$. This technique is now called \emph{domain coloring}. Using
hue and lightness to represent argument and magnitude is the default complex
plotting method in SageMath~\cite{sage}, as well as other common complex
plotting libraries.

In principle, such a graph gives perfect information about the function. But in
practice, it is often difficult to distinguish between hue and lightness
--- and in particular it is challenging to determine if two points with
different hues have the same lightness. Thus some variations omit a dimension
from the graph and plot only the magnitude or phase. Even though the resulting
graphs do not perfectly capture $f$, the omission of a dimension can be
beneficial. Other variations specially color the complex plane in other ways.

In~\cite{wegert2010phase}, Wegert describes an approach using domain coloring,
but emphasizing phase plots as a visual tool. He also produces phase plots with
small brightness adjustments near certain magnitude thresholds. The effect is
similar to looking at a contour map of a landscape, except that each pixel's
color carries meaning (the phase of $f$ evaluated at that point).

\subsection{Paper Overview}

In~\S\ref{sec:survey}, we give a short survey of common existing complex
plotting techniques applied to modular forms. We include images coming from the
ideas of Farris and Wegert, built using SageMath. We also reproduce the plots
associated to modular forms on the L-Function and Modular Form Database
(LMFDB~\cite{lmfdb}).

In~\S\ref{sec:color}, we describe how one can incorporate pre-established
colormaps into modular form visualizations.
And in~\S\ref{sec:implementation},
we discuss one such implementation in terms of \texttt{matplotlib}. This is a
new implementation.

Throughout, we revisit the same basic figures with each visualization
technique. We plot four different modular forms on three different domains in
each technique. Each plot reveals different behaviors of the underlying forms.

\section{Survey of Visualizations}\label{sec:survey}

\subsection{Halfplane and Disk Models}

Plots of modular forms typically represent the form either on the upper
halfplane or on the Poincar\'{e} disk. We first describe the relationship
between these two types of plots, and then survey different visualizations of
forms in each plot type.

Let $\mathcal{H} = \{ x + iy : y > 0 \}$ denote the complex upper halfplane.
A modular form of weight $k$ on a congruence subgroup
$\Gamma \subseteq \SL(2, \mathbb{Z})$ is a holomorphic function $f: \mathcal{H}
\longrightarrow\mathbb{C}$ satisfying
\begin{equation}\label{eq:automorphy}
  f(\gamma z) = {(cz + d)}^k f(z) \qquad \forall \gamma =
  \big(\begin{smallmatrix} a & b \\ c & d \end{smallmatrix}\big) \in \Gamma,
\end{equation}
as well as certain growth conditions at the cusps of $\Gamma
\backslash \mathcal{H}$. See~\cite{diamond2005first} for complete
definitions.

To the congruence subgroup $\Gamma$ is associated a positive integer $N$ called
the level. The automorphy condition~\eqref{eq:automorphy} implies strong
symmetry conditions on $f$ in terms of $N$. In particular,
$\big(\begin{smallmatrix} 1&N \\ 0&1\end{smallmatrix}\big) \in \Gamma$ implies
that $f$ will be periodic with period $N$. Thus one can constrain a plot to a
vertical strip of width $N$ without omitting any information.

In practice, one would cut off the vertical strip at some height $H$.
If $f$ is a cuspform, so that $\lim_{y \to \infty} f(x + iy) = 0$, then
heuristically most interesting behavior of the form will lie within a
well-chosen box $[0, N] \times [0, H]$. In practice, $N$ is typically much
larger than $H$, and it's necessary to investigate some smaller box.
We refer to this type of plot as a plot of $f$ in the upper halfplane.

Alternately, the upper halfplane is conformally equivalent to the Poincar\'{e}
disk $\mathbb{D} = \{ x + i y : x^2 + y^2 < 1 \}$. There are infinitely many
such maps taking $\mathbb{D}$ to $\mathcal{H}$, but we choose the m\"{o}bius
transform
\begin{equation}
  \phi(z) = \frac{1 - iz}{z - i}.
\end{equation}
Under $\phi$, the points $-i, 0, i$ in $\mathbb{D}$ are mapped to $0, i,
i\infty$, respectively, in $\mathcal{H}$. Thus the apparent vertical orientation
remains fixed in both models.

Plotting a modular form $f$ on $\mathbb{D}$ is a complete picture; values of $f$
at every point of $\mathcal{H}$ will be represented through such a plot. We
refer this type of plot as a plot of $f$ in the disk.

\subsection{Visualizations}\label{ssec:visualizations}

Let $\Delta(z)$ denote the Ramanujan Delta function, the unique holomorphic
cuspform of weight $12$ on $\SL(2, \mathbb{Z})$. And let $g(z)$ denote the
unique holomorphic cuspform of weight $4$ on $\Gamma(5)$.
We think of $g$ as a ``simple'' modular form, and $\Delta$ is perhaps the
most widely-recognized modular form.
Each of these forms is catalogued in the LMFDB~\cite{lmfdb}, referenced by a
label. The $\Delta$ function has
label
\texttt{1.12.a.a.1.1}\footnote{\url{https://www.lmfdb.org/ModularForm/GL2/Q/holomorphic/1/12/a/a/1/1/}}
and $g$ has label
\texttt{5.4.a.a.1.1}\footnote{\url{https://www.lmfdb.org/ModularForm/GL2/Q/holomorphic/5/4/a/a/1/1/}}.

We also consider two more complicated modular forms:
$f_{105}$, a weight $2$ modular form on
$\Gamma(105)$, identified by label \texttt{105.2.a.a.1.1}\footnote{\url{https://www.lmfdb.org/ModularForm/GL2/Q/holomorphic/105/2/a/a/1/1/}}; and
$f_{10}(z)$, a weight $20$ modular form on $\Gamma(10)$, identified by
label
\texttt{10.20.a.a.1.1}\footnote{\url{https://www.lmfdb.org/ModularForm/GL2/Q/holomorphic/10/20/a/a/1/1/}}.

We should expect forms of similar weight to share characteristics, and we
should expect forms whose levels have similar numbers of cusps to share some
characteristics. Here, we've chosen levels that correlate strongly with the
number of cusps. Thus the two forms of smaller level have fewer distinct cusps
than the two forms of larger level.

Note that $g(z)$ has small level and small weight, $\Delta(z)$ has small level
and moderately large weight, $f_{105}$ has very large level and very small
weight, and $f_{10}$ has moderately large level and high weight,

In the plots that follow, we will show these four forms in the order $(g,
\Delta, f_{105}, f_{10})$ from top to bottom. To produce the
plots, we approximate each modular form by its first $400$ Fourier coefficients
and evaluate the function on a $600 \times 600$ grid, and store it as a PNG.
We plot each form on the Poincar\'e disk, on a natural box in $\mathcal{H}$,
and then on a ``zoomed-in'' portion of $\mathcal{H}$. Specifically, all
``zoomed-in'' plots on $\mathcal{H}$ are of the region $[0.1, 0.4] \times [0,
0.25]$. For the other plots on $\mathcal{H}$, we plot $\Delta$ on $[-1, 1]
\times [0, 2]$, $g$ on $[-2.5,
2.5] \times [0, 2]$, $f_{105}$ on
$[-1, 1] \times [0, 1]$, and $f_{10}$ on $[-1, 1] \times [0, 2]$.

The region choices for $\Delta$ and $g$ are natural: these are each wide enough
to include at least one period, and tall enough to capture most interesting
properties. (This will be apparent in the graphs). For $f_{105}$ and $f_{10}$,
it's not practical to plot a box sufficiently wide to capture a period.
Instead we plot a smaller region.

We plot each of these four forms on each of the three regions with a variety of
different visualizations and discuss each in turn.

\begin{remark}

We do not intend to emphasize one type of visualization as being universally
better than all the others. As we will see, each type has flaws and only
presents a single view. Nonetheless, at the end of paper we give our
recommendations for producing visualizations.

\end{remark}

\subsubsection{Standard domain coloring}\label{sec:standard_domain_coloring}

\begin{figure}[!t]
\centering
\includegraphics[width=0.32\textwidth]{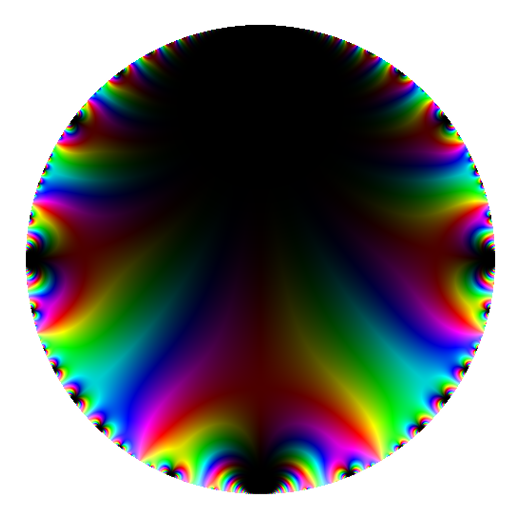}
\raisebox{0.5in}{\includegraphics[width=0.32\textwidth]{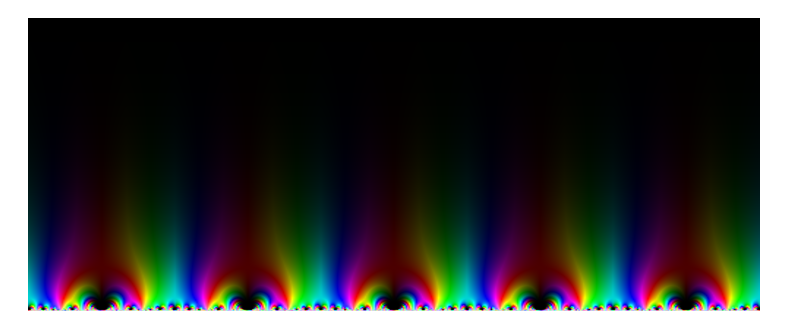}}
\raisebox{0.5in}{\includegraphics[width=0.32\textwidth]{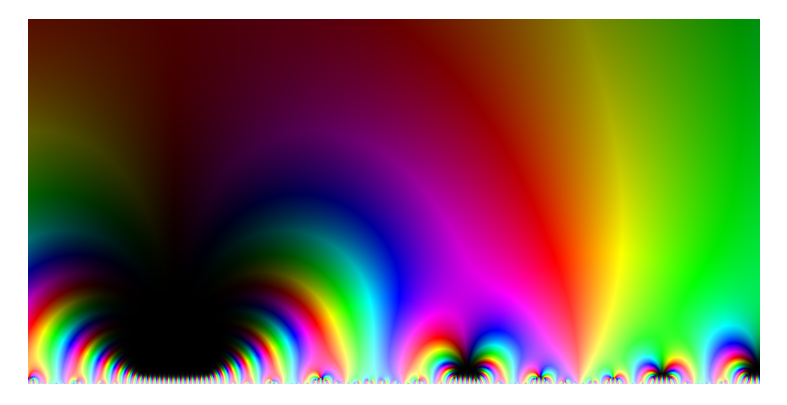}}

\includegraphics[width=0.32\textwidth]{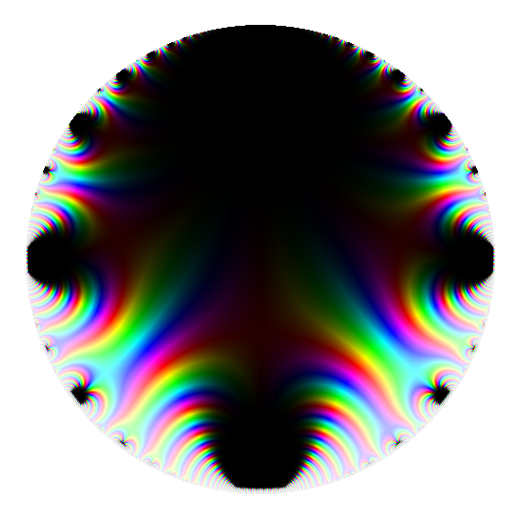}
\includegraphics[width=0.32\textwidth]{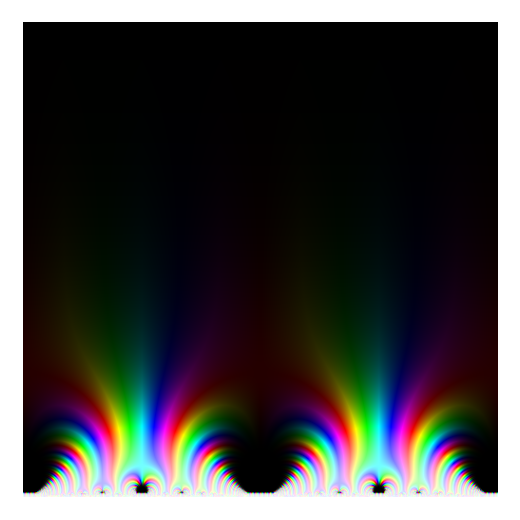}
\raisebox{0.5in}{\includegraphics[width=0.32\textwidth]{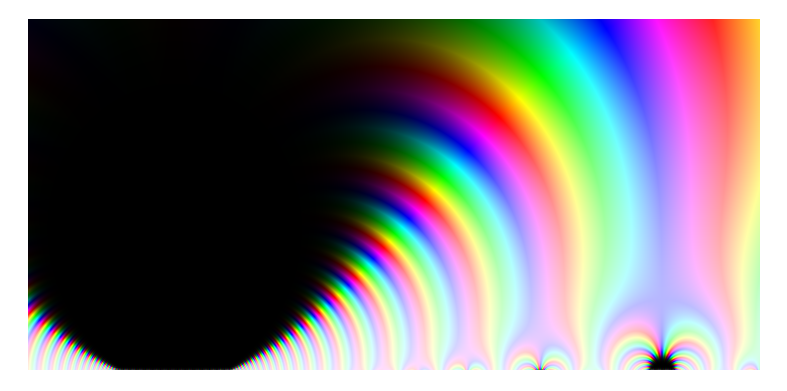}}

\includegraphics[width=0.32\textwidth]{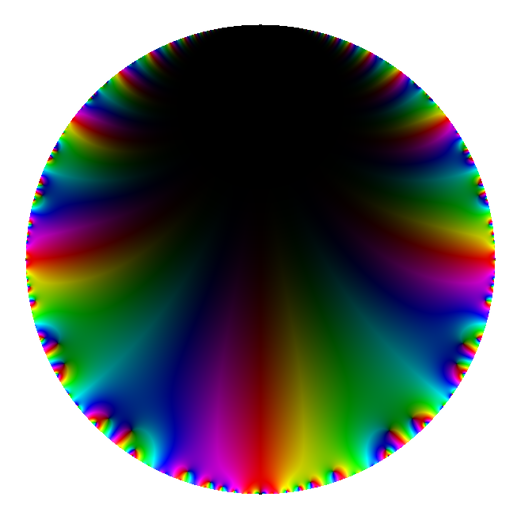}
\raisebox{0.5in}{\includegraphics[width=0.32\textwidth]{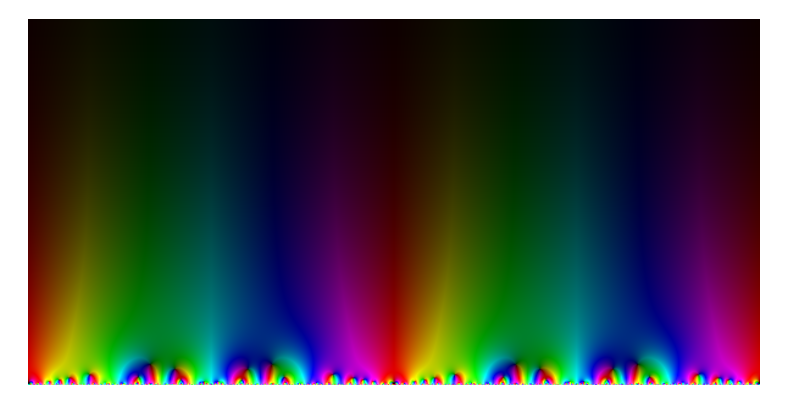}}
\raisebox{0.5in}{\includegraphics[width=0.32\textwidth]{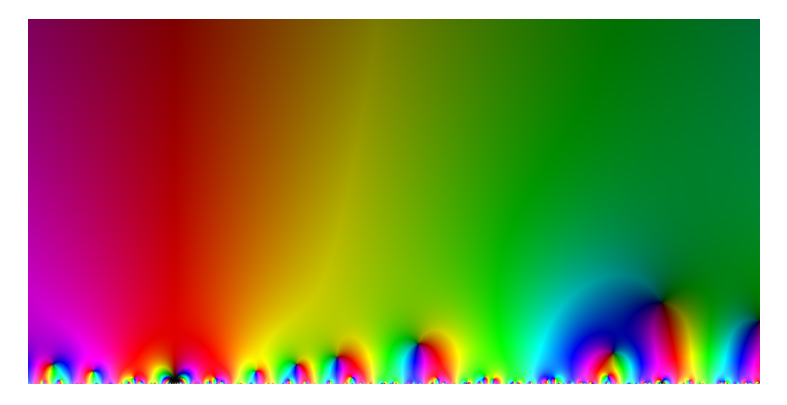}}

\includegraphics[width=0.32\textwidth]{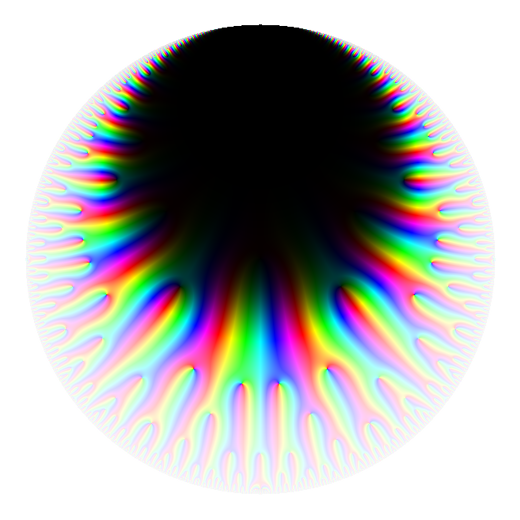}
\includegraphics[width=0.32\textwidth]{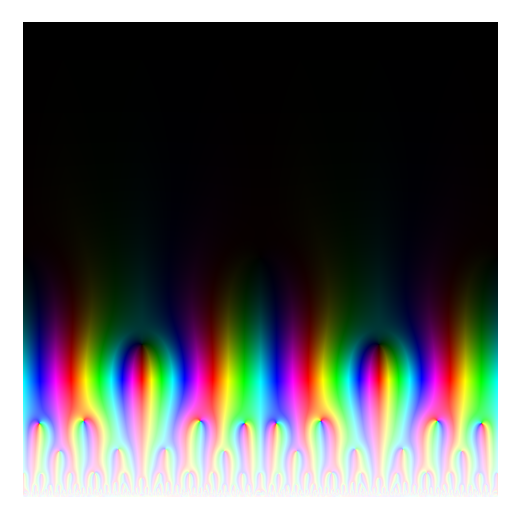}
\raisebox{0.5in}{\includegraphics[width=0.32\textwidth]{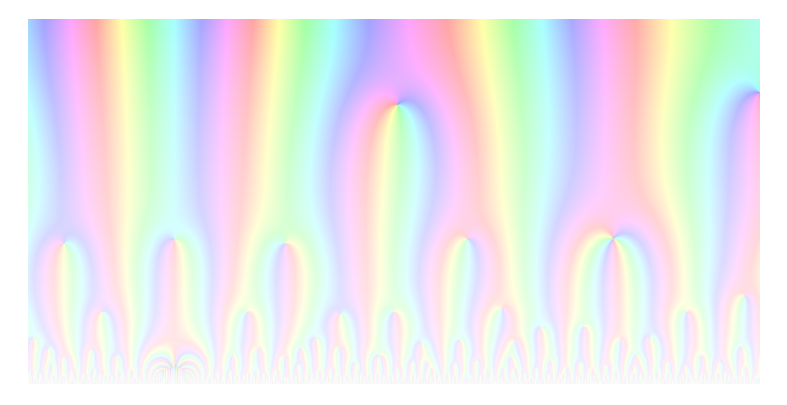}}

\caption{Default plots using \texttt{complex\_plot} in SageMath. From top to
bottom, the forms are $g, \Delta, f_{105}$, and $f_{10}$.}\label{fig:SDC_1}
\end{figure}

In Figure~\ref{fig:SDC_1}, we present default plots (as created by
\texttt{complex\_plot} in SageMath). In these plots, the $\arg z$ is represented
by hue and $\lvert z \rvert$ is represented by brightness, where $0$ is black
and $\infty$ is white.

The fact that each form plotted is a cuspform is immediately clear as each plot
on the disk is dominated by the color black (corresponding to the overall size
being small). Further, a closer look will reveal that each cusp appears to be
visibly dark. This is most obvious in the two lower level forms, $g$ and
$\Delta$, where each cusp has a relatively large dark region around it. As the
number of cusps increases, the behavior around cusps becomes more varied. Very
near the boundary, our approximation for the form is less accurate. This is one
factor contributing to why the cusps of $f_{10}$ are far less pronounced.

The weight of each form manifests primarily in how quickly the form can go from
very bright to very dark. Here, the two forms of higher weight, $\Delta$ and
$f_{10}$, have regions of near-white brightness near the boundary that very
rapidly darken towards the top of the disk. The two forms of lower weight are
much more even in brightness. The high weight of $f_{10}$ is another factor
contributing to why its cusps are less pronounced: there are smaller regions of
darkness around each cusp.

It is possible to adjust the mapping from magnitude to brightness, but in
practice these plots look roughly the same for a wide choice of maps and it
takes quite a bit of tuning to choose an appropriate map.\footnote{Adjusting
the map from magnitude to brightness for SageMath's \texttt{complex\_plot} is
possible, but nontrivial; the map is defined internally and the current
plotting interface doesn't give any option to choose or alter this map. In
order to adjust this map, it is necessary to modify the source for SageMath's
plotting routines directly.}

\subsubsection{Magnitude plot without color}\label{sec:mag_no_color}

\begin{figure}[!t]
\centering
\includegraphics[width=0.32\textwidth]{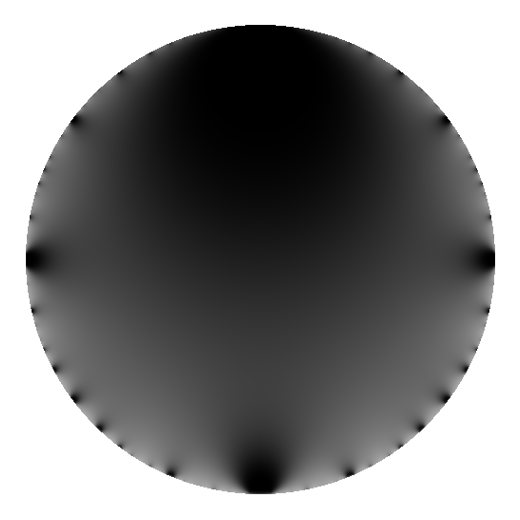}
\raisebox{0.5in}{\includegraphics[width=0.32\textwidth]{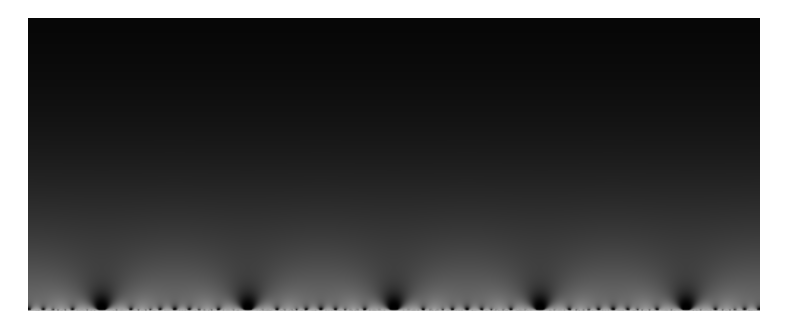}}
\raisebox{0.5in}{\includegraphics[width=0.32\textwidth]{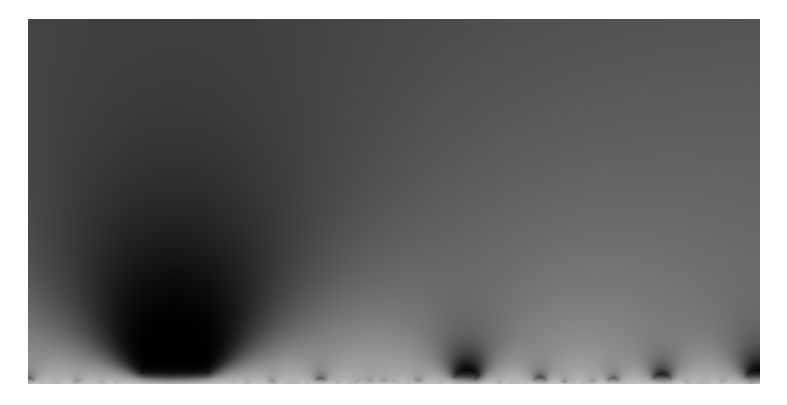}}
\includegraphics[width=0.32\textwidth]{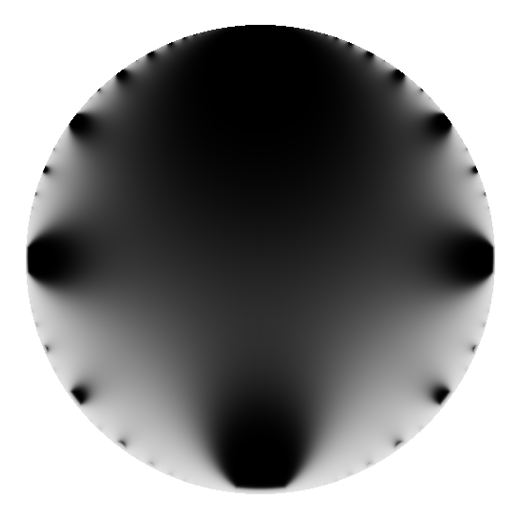}
\includegraphics[width=0.32\textwidth]{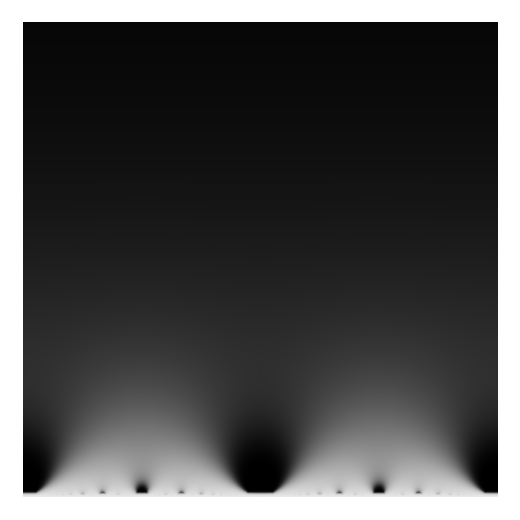}
\raisebox{0.5in}{\includegraphics[width=0.32\textwidth]{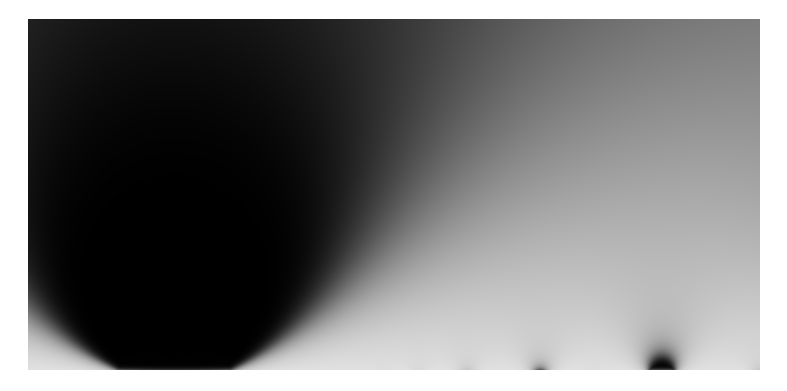}}
\includegraphics[width=0.32\textwidth]{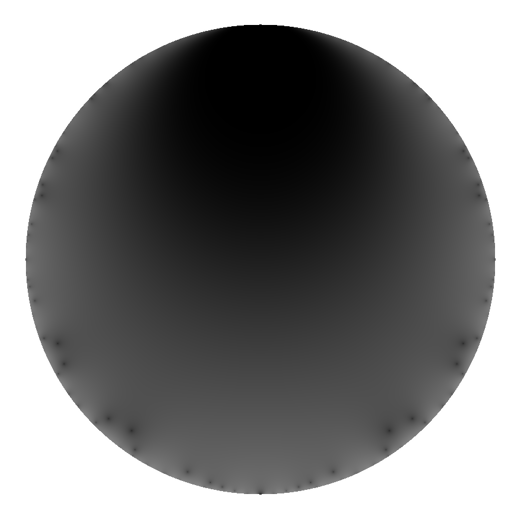}
\raisebox{0.45in}{\includegraphics[width=0.32\textwidth]{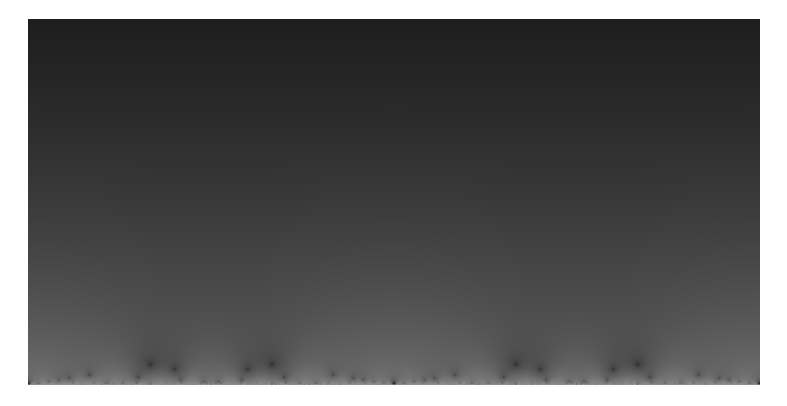}}
\raisebox{0.45in}{\includegraphics[width=0.32\textwidth]{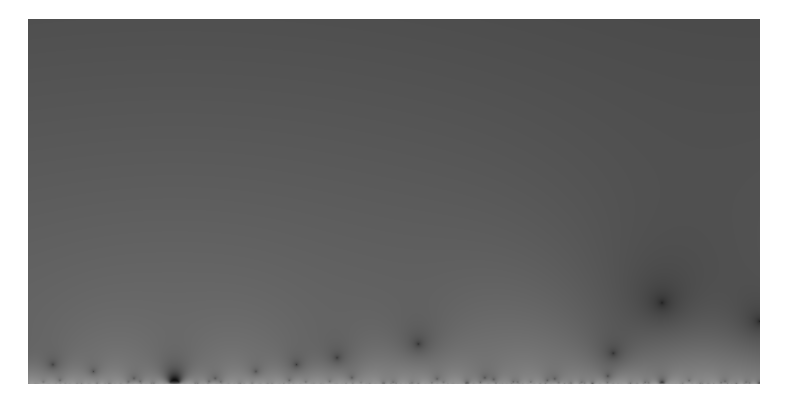}}
\includegraphics[width=0.32\textwidth]{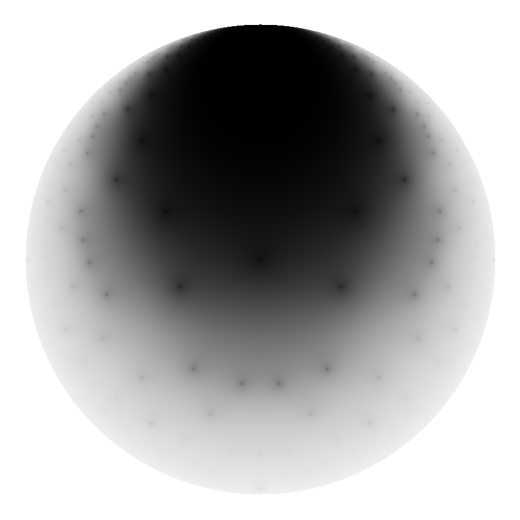}
\includegraphics[width=0.32\textwidth]{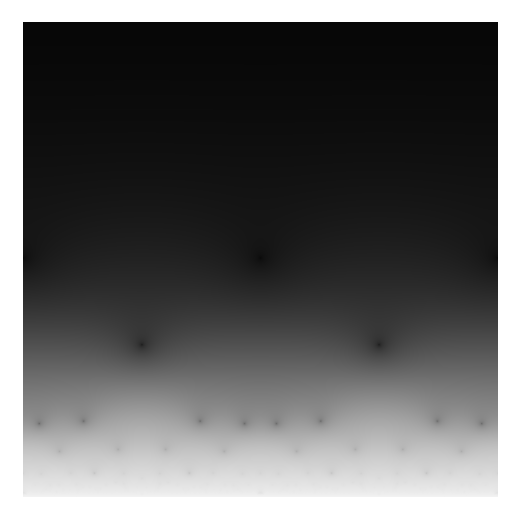}
\raisebox{0.5in}{\includegraphics[width=0.32\textwidth]{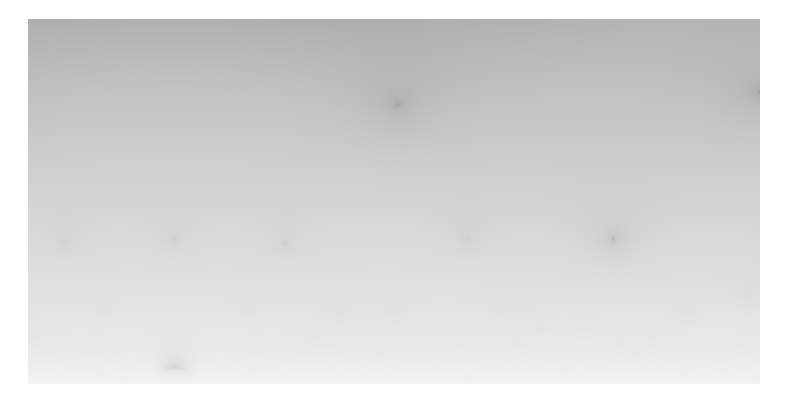}}
\caption{Magnitude plots where the magnitude corresponds to darkness. Black
corresponds to zero magnitude. From top to bottom, the forms are $g, \Delta,
f_{105}$, and $f_{10}$.}\label{fig:grey}
\end{figure}

We now produce plots after ignoring the phase of $f(z)$ and instead only
using the magnitude $\lvert f(z) \rvert$. There remains a choice of how to map
magnitude to brightness; it is common to use maps of the form
$\arctan(\log({\lvert f(z) \rvert}^\alpha + 1))$ for some $0 < \alpha < 1$.
Taking $\alpha = 0.25$ yields the plots in Figure~\ref{fig:grey}. We note that
adjusting the power $\alpha$ is the most natural and clearest way to adjust
the overall darkness/lightness scale of this style of plot: larger alpha causes
points near zeros to be much darker and points near poles to be much lighter;
smaller alpha decreases this effect.

Qualitatively, each of these pictures offers very similar information as those
in Figure~\ref{fig:SDC_1}. The lack of phase information makes it slightly more
challenging to distinguish between the images of $\Delta$ and $g$ in the disk.
But note that it is very easy to determine forms of low weight and high weight,
as the forms with low weight have much slower changes in brightness.

The complicated and varied behavior of $f_{10}$ near the boundary is more
evident in this plot. The presence of many more points of small size is made
far more obvious. It is just barely possible to observe the region of
decreasing size around the $0$ cusp.

\subsubsection{Magnitude plot with periodic linear
color}\label{sec:mag_periodic_lin}

\begin{figure}[!h]
\centering
\includegraphics[width=0.32\textwidth]{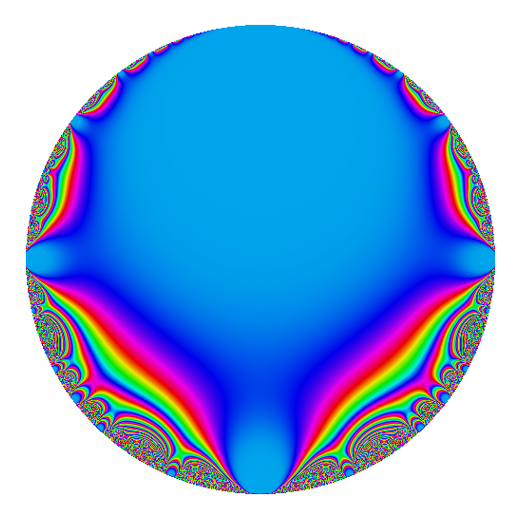}
\raisebox{0.5in}{\includegraphics[width=0.32\textwidth]{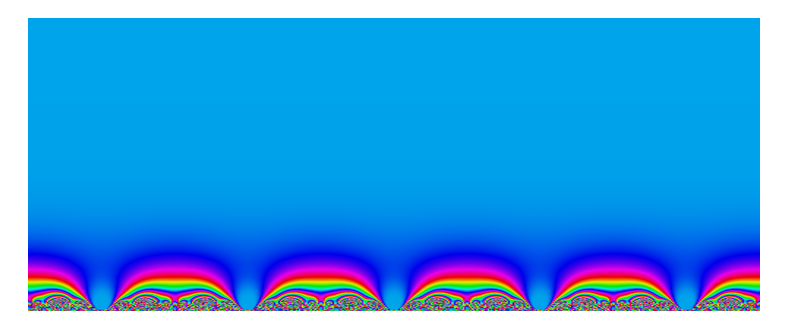}}
\raisebox{0.5in}{\includegraphics[width=0.32\textwidth]{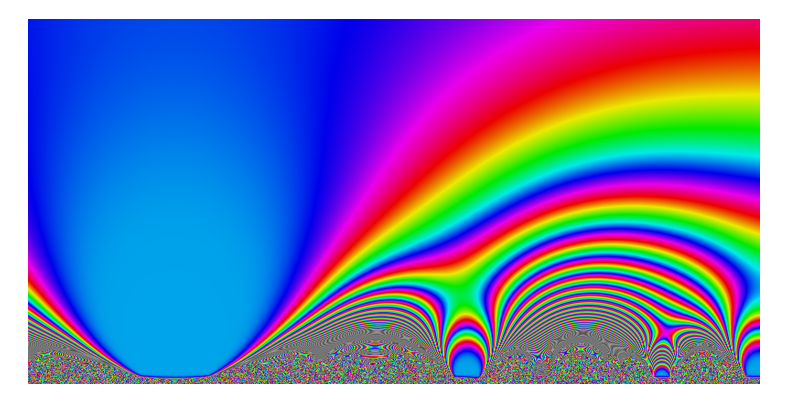}}
\includegraphics[width=0.32\textwidth]{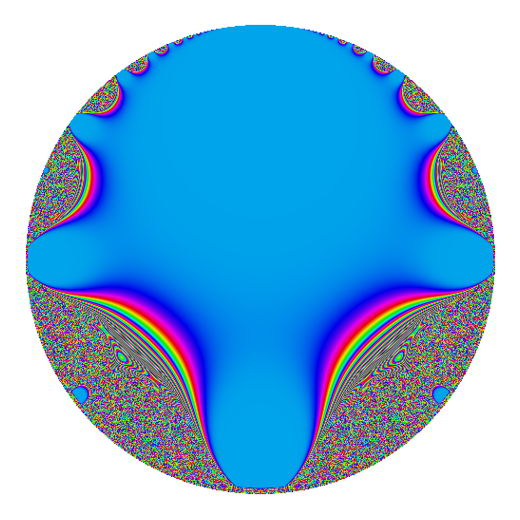}
\includegraphics[width=0.32\textwidth]{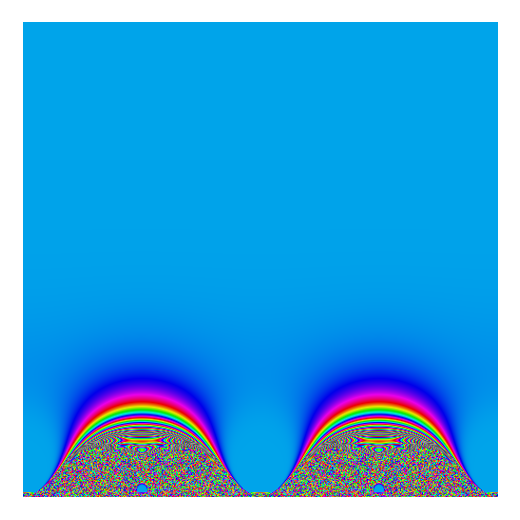}
\raisebox{0.5in}{\includegraphics[width=0.32\textwidth]{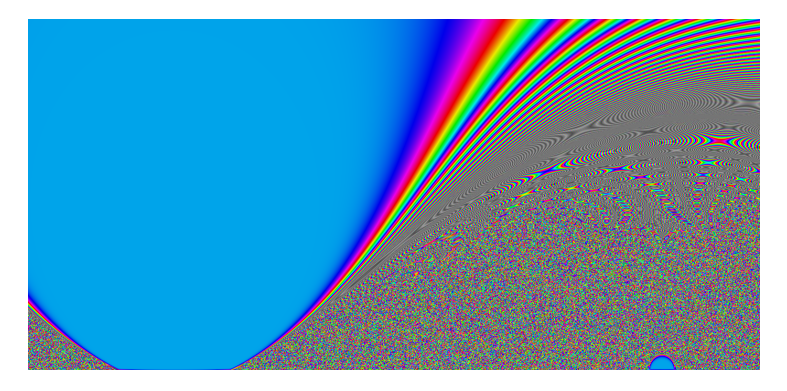}}
\includegraphics[width=0.32\textwidth]{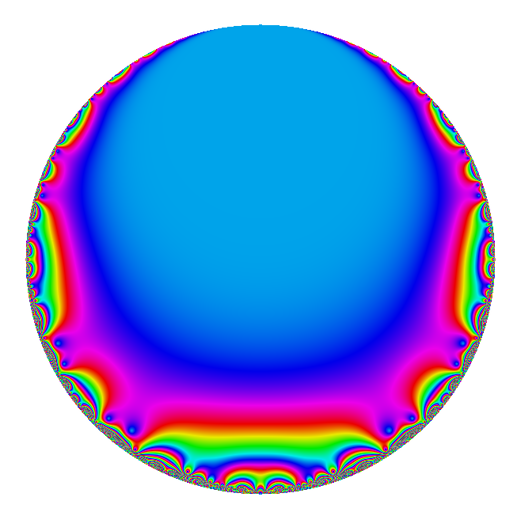}
\raisebox{0.45in}{\includegraphics[width=0.32\textwidth]{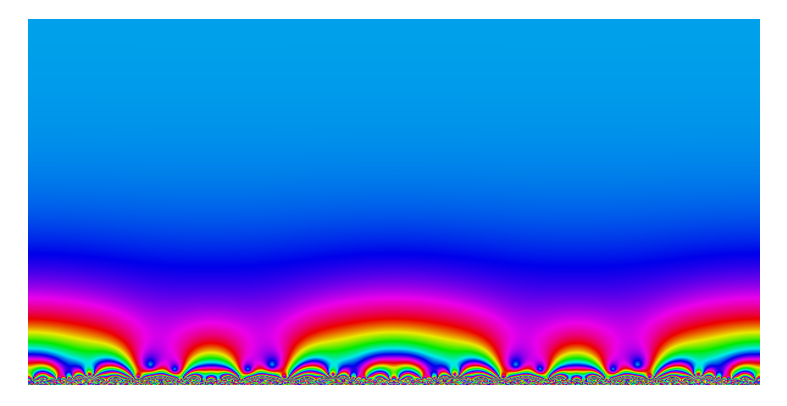}}
\raisebox{0.45in}{\includegraphics[width=0.32\textwidth]{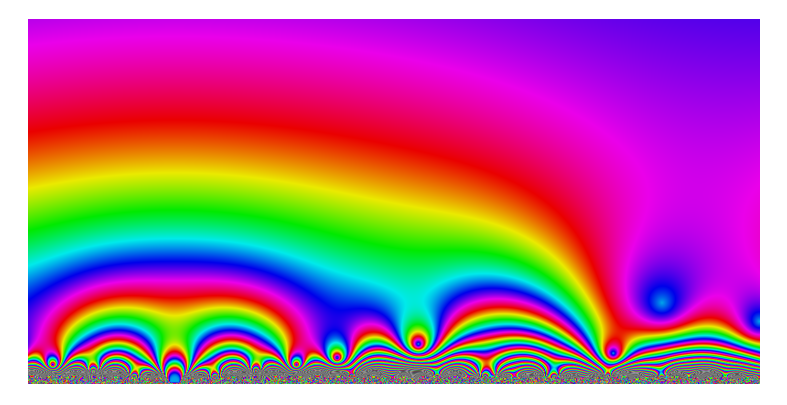}}
\includegraphics[width=0.32\textwidth]{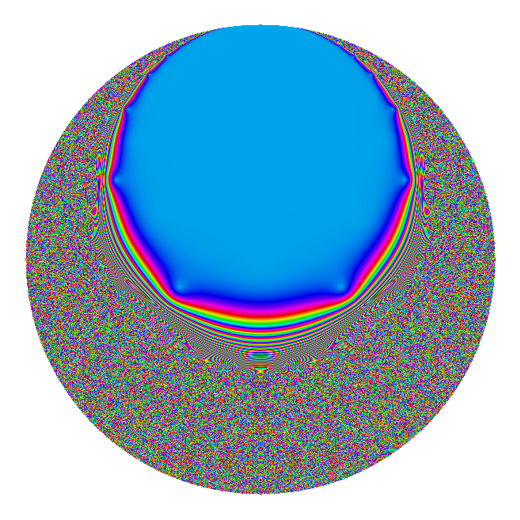}
\includegraphics[width=0.32\textwidth]{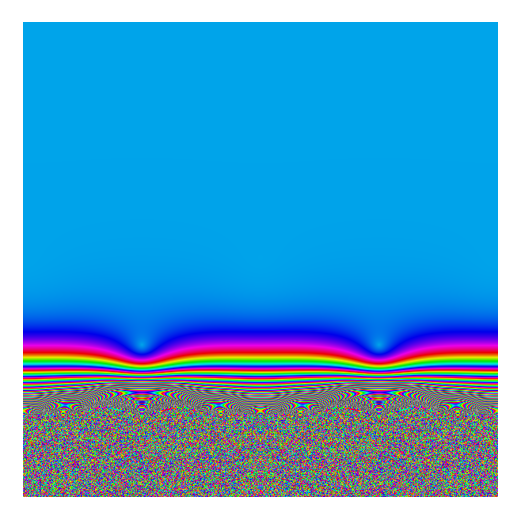}
\raisebox{0.5in}{\includegraphics[width=0.32\textwidth]{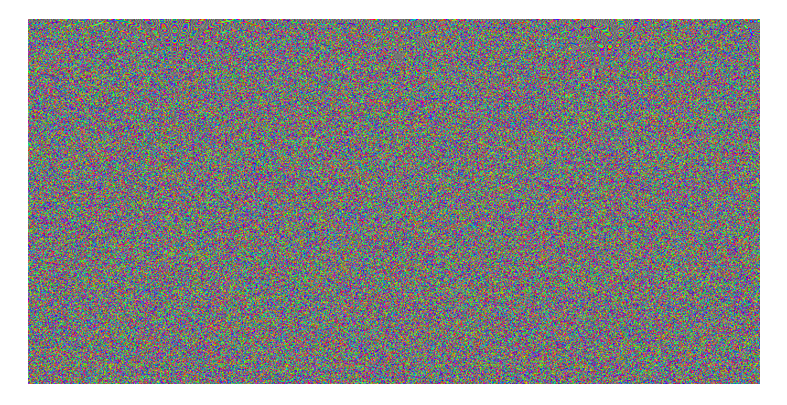}}
\caption{Magnitude plots mimicking the style of the LMFDB.\@ The phase is
ignored; the hue represents the magnitude modulo $1$. From top to bottom, the
forms are $g, \Delta, f_{105}$, and $f_{10}$.}\label{fig:LMFDB}
\end{figure}

In the LMFDB, magnitude is mapped periodically to color.
That is, hue represents the magnitude modulo $1$, where blue is zero, and
increases through purple, red, orange, yellow, and so on.
In Figure~\ref{fig:LMFDB}, we mimic the plots in the LMFDB attained by ignoring
the phase of mapping the magnitude mod $1$ to hue.

Recall from \S\ref{ssec:visualizations} that each of the four forms we are
considering has a label on the LMFDB, and we gave links to their homepages.
Each page has a medallion with a plot similar to the plots on the Poincar\'e
disk given here.

The large bluish blob corresponds to a region where the magnitude of the form
is never more than $1$. The larger weights of $\Delta(z)$ and $f_{10}$
correspond to more rapid growth away from cusps, leading both to smaller bluish
regions and more tightly packed lines; $f_{10}$ has larger weight, and both of
these traits are even more prominent in its graphs.

For the lower weight forms $g$ and $f_{105}$, this is a very accurate plot. But
$\Delta(z)$ and $f_{10}$ change too rapidly near the boundary, causing a
static-like appearance. Increasing resolution helps, but in practice it is not
feasible to produce sufficiently high resolution images to completely make
sense of the boundary with this type of plot.

This sort of plot is very well-suited for low weight forms of any level. The
small weight means that even at moderate resolution, the image will appear
well-defined even near the boundary. Then the cusps associated to the level
will be visually apparent by the varying types of behaviors near the
boundary. This is most obvious in the detail images for $g$ and $f_{105}$,
where we can clearly see the greater variety in behavior near the cusps at the
boundary in plots of $f_{105}$.

\subsubsection{Magnitude plot with periodic logarithmic
spacing}\label{sec:mag_periodic_log}

\begin{figure}[!ht]
\centering
\includegraphics[width=0.32\textwidth]{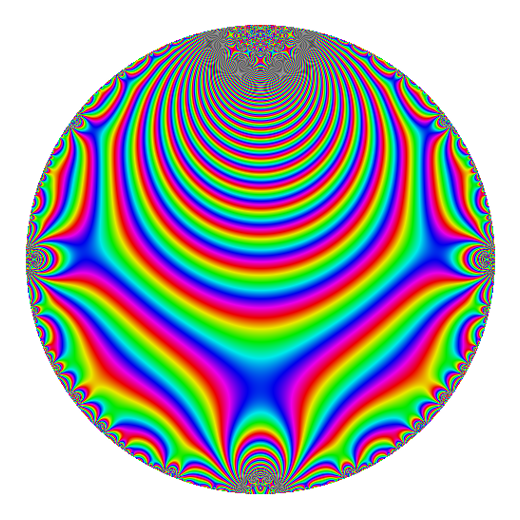}
\raisebox{0.5in}{\includegraphics[width=0.32\textwidth]{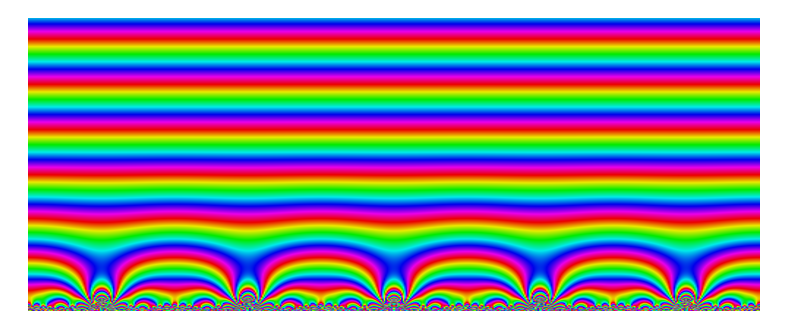}}
\raisebox{0.5in}{\includegraphics[width=0.32\textwidth]{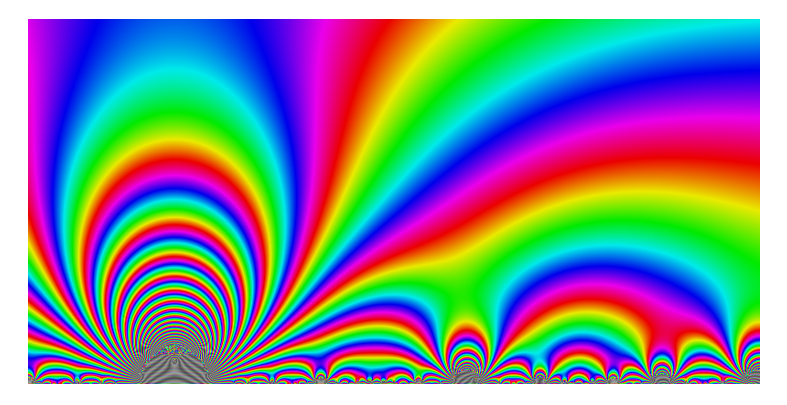}}
\includegraphics[width=0.32\textwidth]{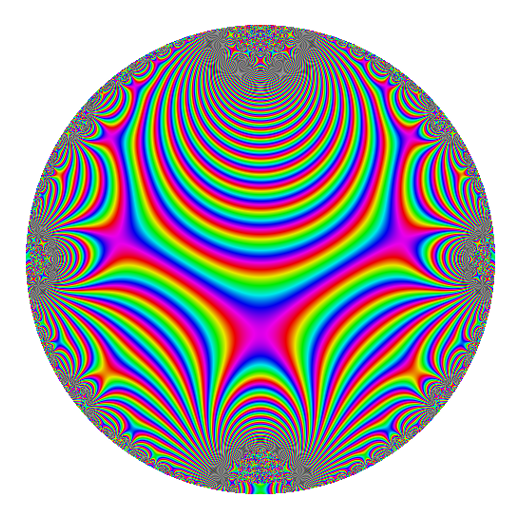}
\includegraphics[width=0.32\textwidth]{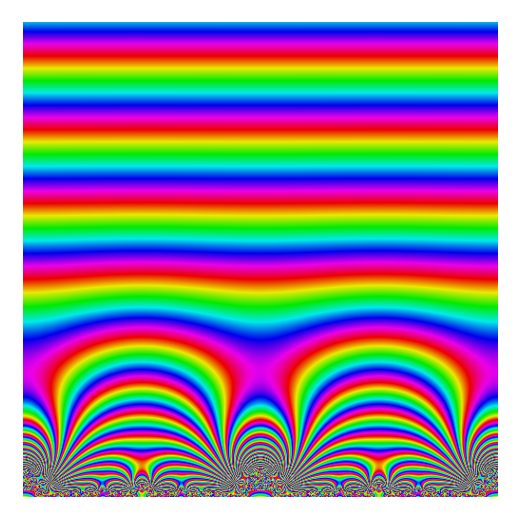}
\raisebox{0.5in}{\includegraphics[width=0.32\textwidth]{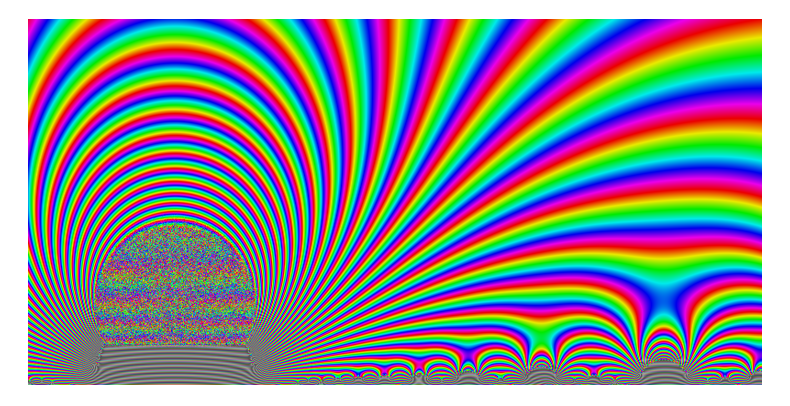}}
\includegraphics[width=0.32\textwidth]{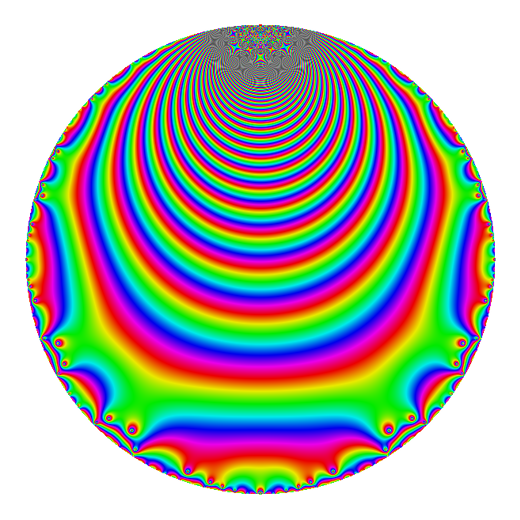}
\raisebox{0.45in}{\includegraphics[width=0.32\textwidth]{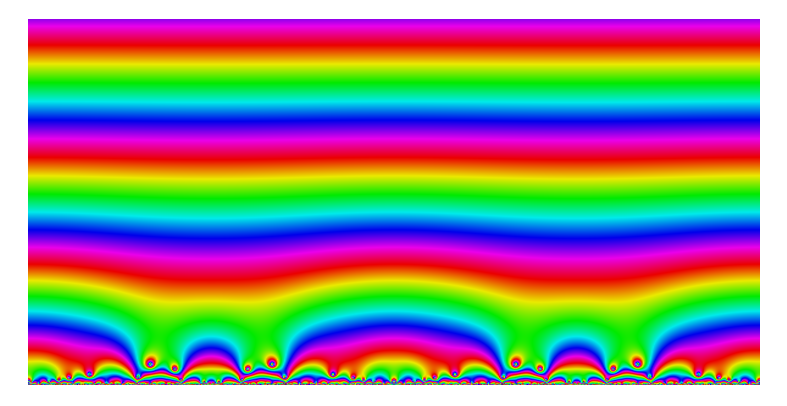}}
\raisebox{0.45in}{\includegraphics[width=0.32\textwidth]{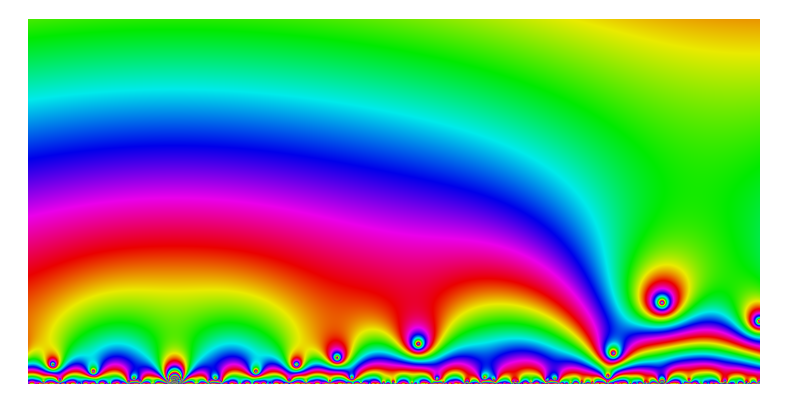}}
\includegraphics[width=0.32\textwidth]{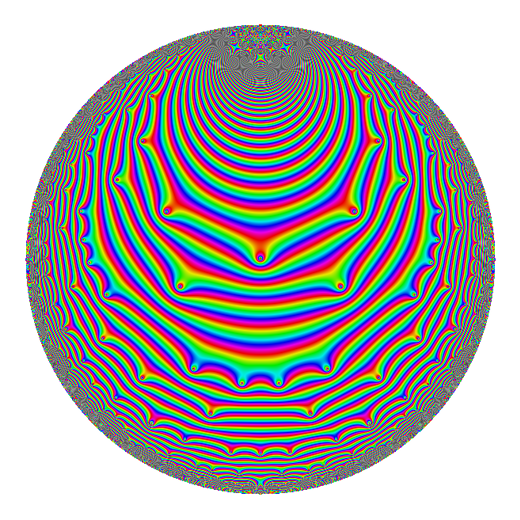}
\includegraphics[width=0.32\textwidth]{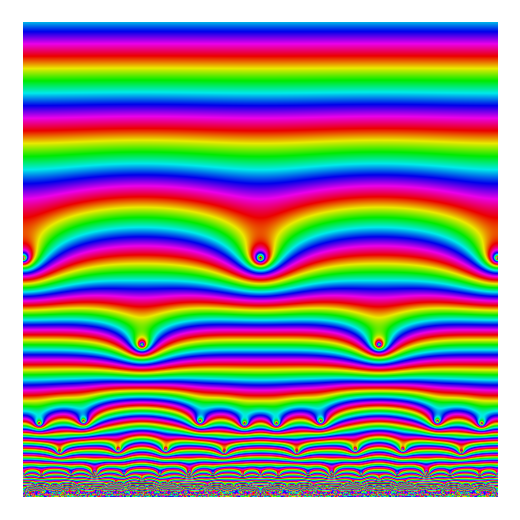}
\raisebox{0.5in}{\includegraphics[width=0.32\textwidth]{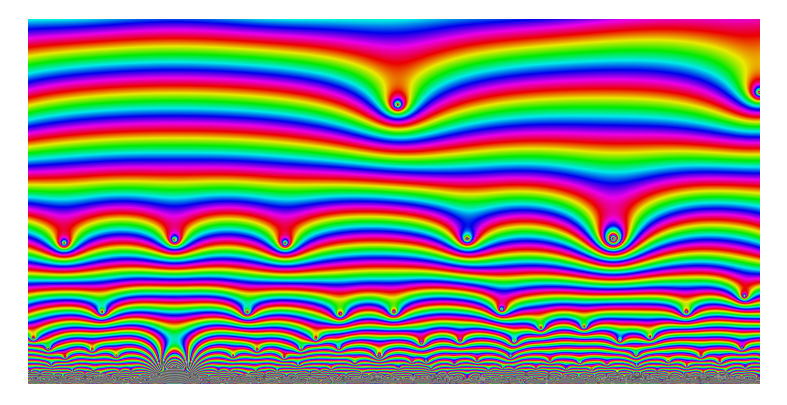}}
\caption{Magnitude plots. The phase is ignored; hue represents the log of the
magnitude, modulo $1$. From top to bottom, the forms are $g, \Delta, f_{105}$,
and $f_{10}$.}\label{fig:LMFDB_log}
\end{figure}

In Figure~\ref{fig:LMFDB_log}, we change yet again the map from magnitude to
color. We now allow the hue to represent $\log_\alpha(\lvert f(z) \rvert) \bmod
1$ for some base $\alpha$. Thus two consecutive bands of red will correspond to
the points in one band being $\alpha$ times the size of the points in the other.

We note that adjusting the factor between consecutive bands (i.e.\ choosing the
base of the logarithm in the map from magnitude to color) controls the number of
spacing of bands, and thus it is possible to make the plots visually simpler or
more complex by experimenting with this parameter. The plots in
Figure~\ref{fig:LMFDB_log} were produced with $\alpha = 7$.

The effect in these plots is similar to giving a higher resolution of
Figure~\ref{fig:LMFDB}. A large amount of detail and data is visible within
these plots; however this is double-edged, and the plots are now busier. The
high growth rate in the forms of higher weight is indicated by many more
tightly packed bands of color near the boundary.

Note that there is a region where the resolution appears insufficient in the
detail plot for $\Delta(z)$. This is actually caused by small errors in the
approximation for $\Delta(z)$ interfering with the extremely tiny values of
$\Delta(z)$ there. As can be seen from the nearly black portions of
Figure~\ref{fig:SDC_1} and Figure~\ref{fig:grey}, $\Delta(z)$ is extremely
small in this region, and thus even small errors in the approximation produce
noticeable artifacts in the graph.

These plots are particularly useful for plots on $\mathcal{H}$, where even for
high weight the plots appear mostly well-formed. In the disk, the
rapid decay at the top of the disk (at the $\infty$ cusp) causes tight
concentric circles of color and attracts a lot of visual attention, even though
the actual behavior is mostly regular and relatively uninteresting there.

\subsubsection{Pure phase plots}\label{sec:pure_phase}

\begin{figure}[!ht]
\centering
\includegraphics[width=0.32\textwidth]{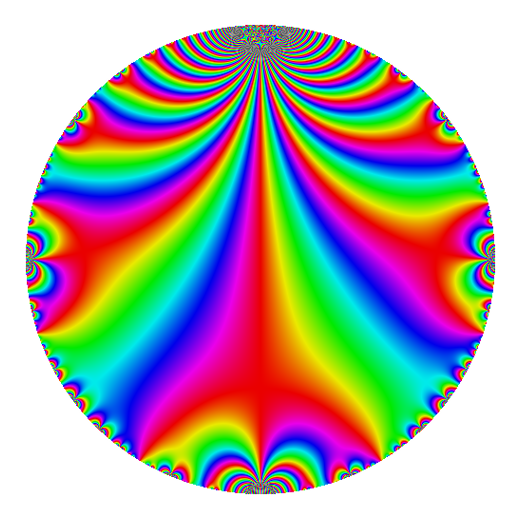}
\raisebox{0.5in}{\includegraphics[width=0.32\textwidth]{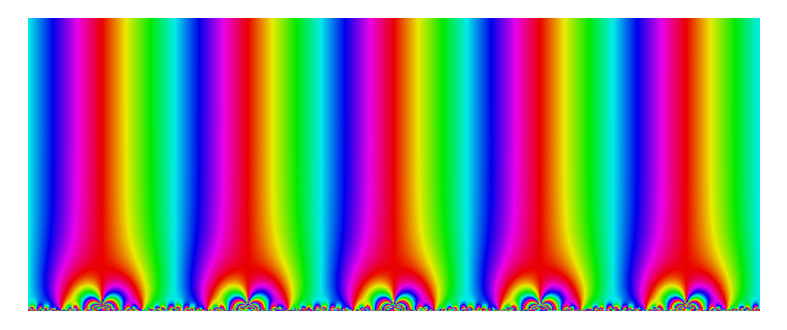}}
\raisebox{0.5in}{\includegraphics[width=0.32\textwidth]{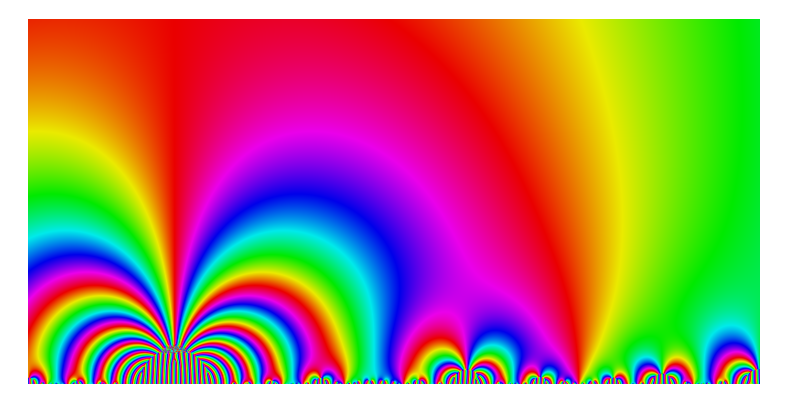}}
\includegraphics[width=0.32\textwidth]{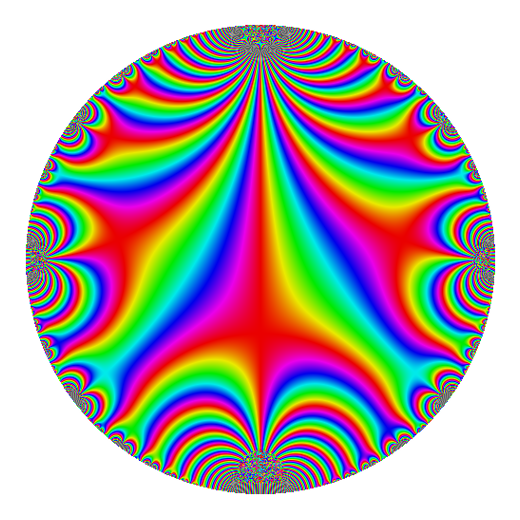}
\includegraphics[width=0.32\textwidth]{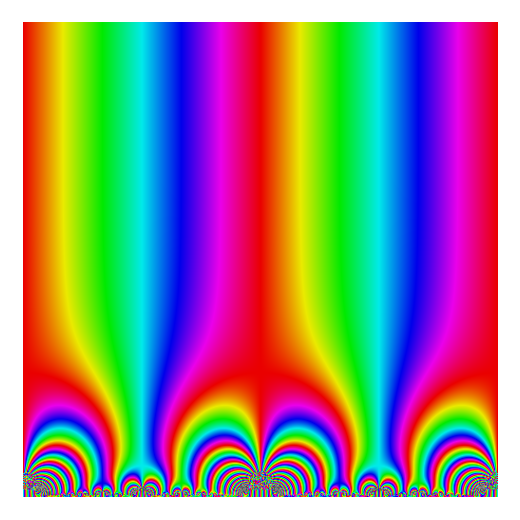}
\raisebox{0.5in}{\includegraphics[width=0.32\textwidth]{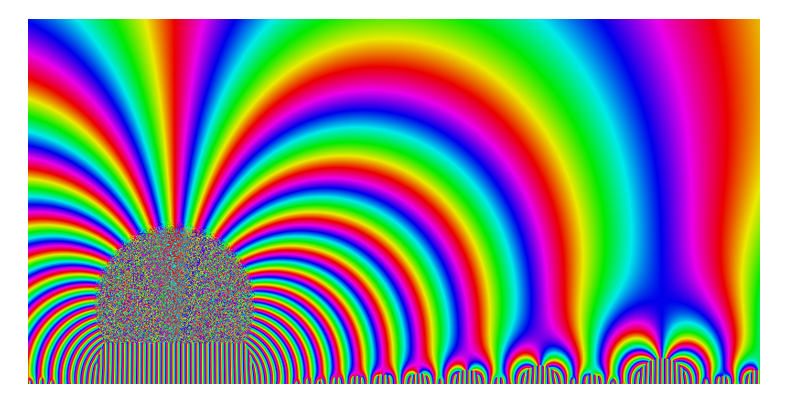}}
\includegraphics[width=0.32\textwidth]{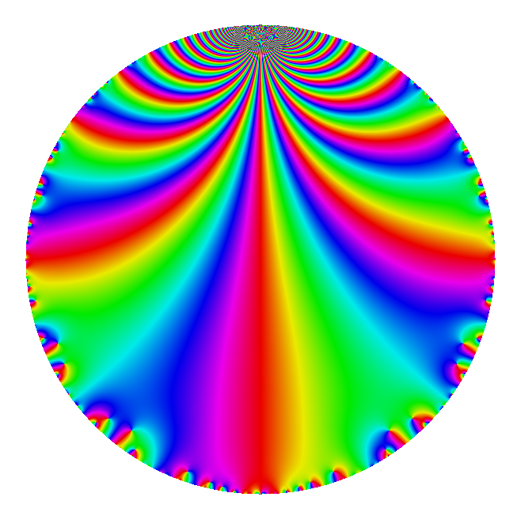}
\raisebox{0.45in}{\includegraphics[width=0.32\textwidth]{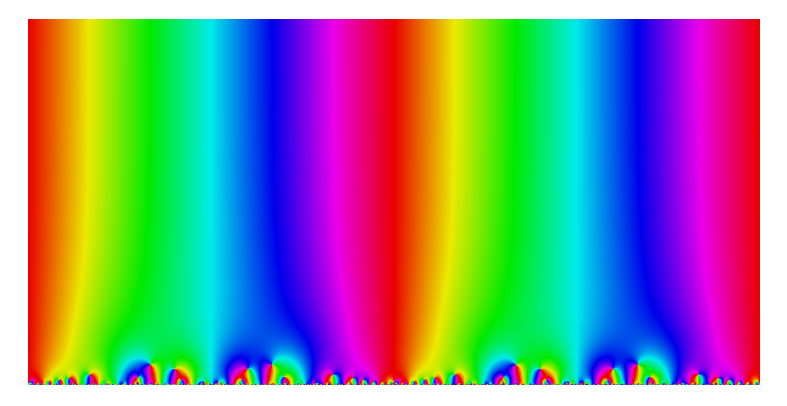}}
\raisebox{0.45in}{\includegraphics[width=0.32\textwidth]{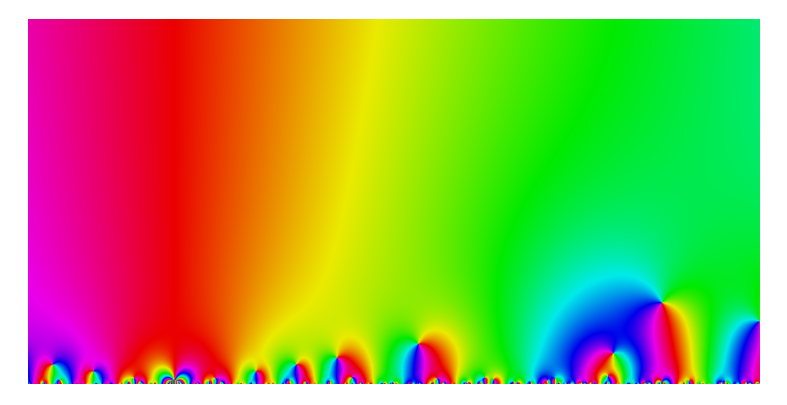}}
\includegraphics[width=0.32\textwidth]{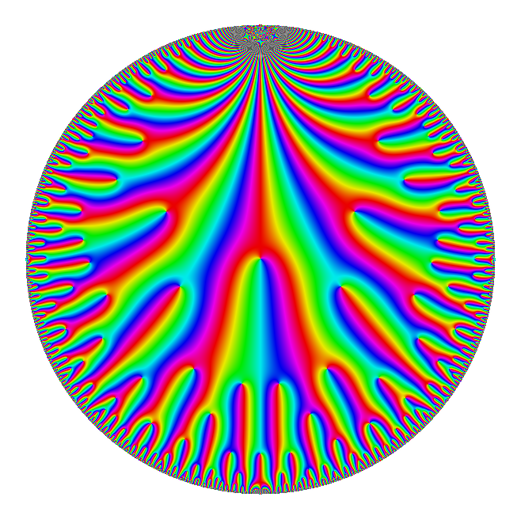}
\includegraphics[width=0.32\textwidth]{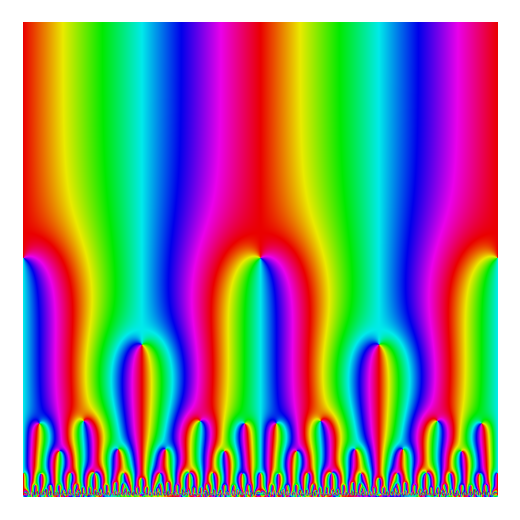}
\raisebox{0.5in}{\includegraphics[width=0.32\textwidth]{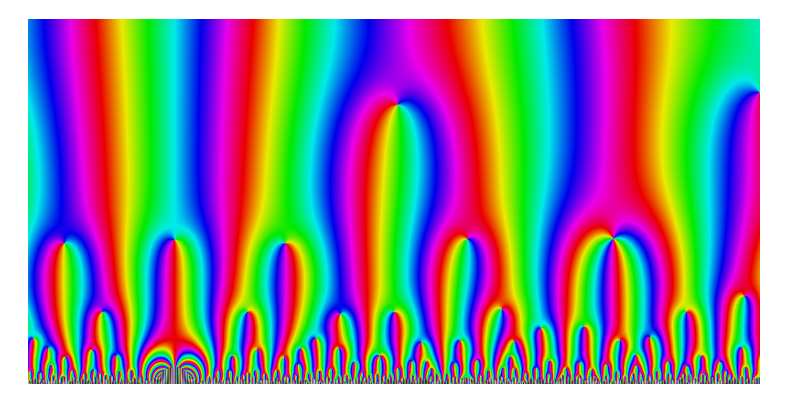}}
\caption{Phase plots. Magnitude is ignored, and hue represents the phase. From
top to bottom, the forms are $g, \Delta, f_{105}$, and
$f_{10}$.}\label{fig:phase}
\end{figure}

Following the ideas of Wegert~\cite{wegert2010phase}, we now consider phase
plots. In Figure~\ref{fig:phase}, we ignore magnitude completely and only
consider the phase. There is a natural mapping from phase to hue: we allow hue
to represent $\arg(z) / 2\pi$, and the natural periodicity of phase works well
with the natural periodicity of the color wheel.

Qualitatively, less behavior of each modular form is evident in its phase plot
in comparison to its magnitude plot. This is partly due to the lack of poles and
infrequent zeros away from the boundary. Poles and zeros are singularities in
phase plots, and the behavior of the colors around these points describes the
order and nature of these points.

In our opinion, pure phase plots should not be used on their own to study or
describe modular forms. However, pairing each image in
Figure~\ref{fig:phase} with its corresponding image in Figure~\ref{fig:LMFDB}
or Figure~\ref{fig:LMFDB_log} completely represents the modular form on the
indicated domain.

\subsubsection{Phase plots with magnitude contours}\label{sec:phase_mag}

\begin{figure}[!ht]
\centering
\includegraphics[width=0.32\textwidth]{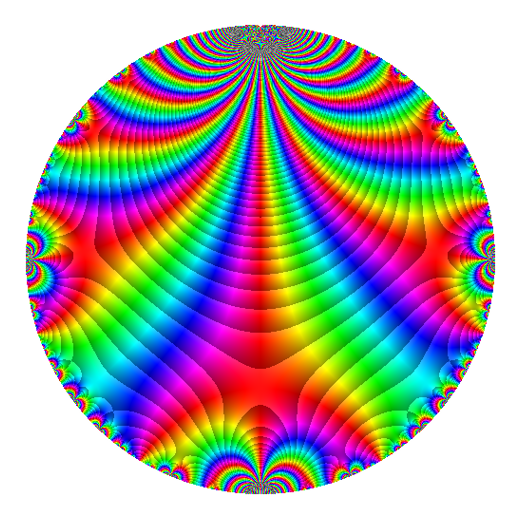}
\raisebox{0.5in}{\includegraphics[width=0.32\textwidth]{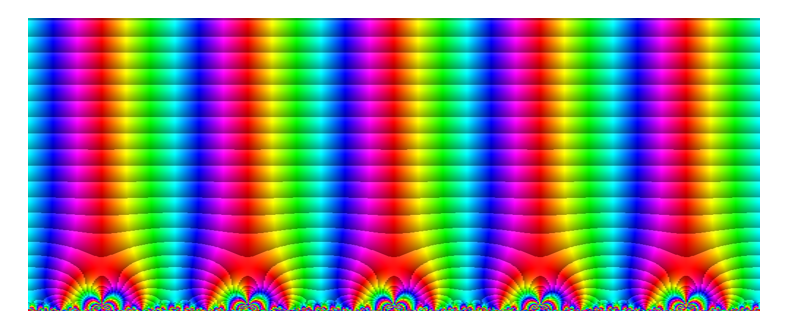}}
\raisebox{0.5in}{\includegraphics[width=0.32\textwidth]{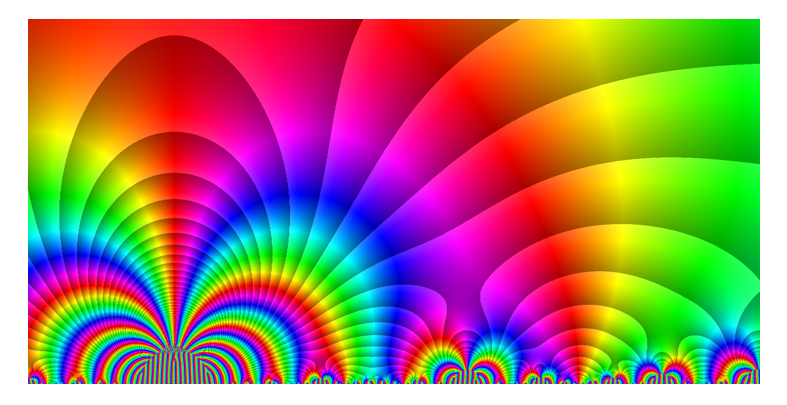}}
\includegraphics[width=0.32\textwidth]{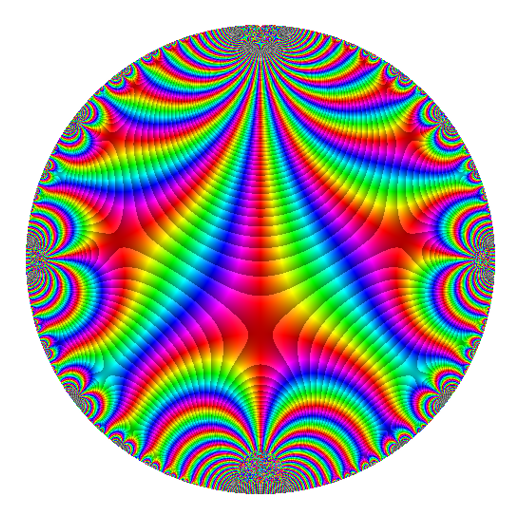}
\includegraphics[width=0.32\textwidth]{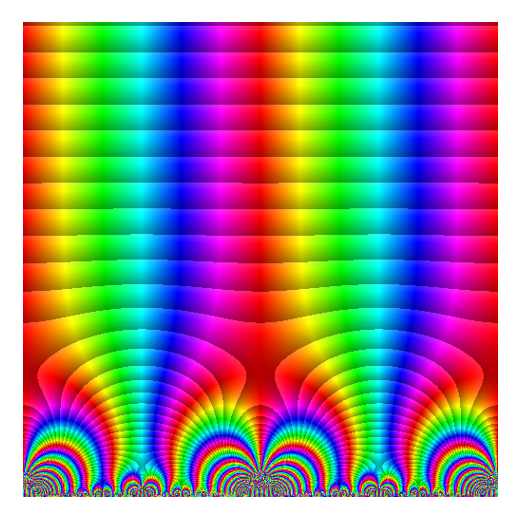}
\raisebox{0.5in}{\includegraphics[width=0.32\textwidth]{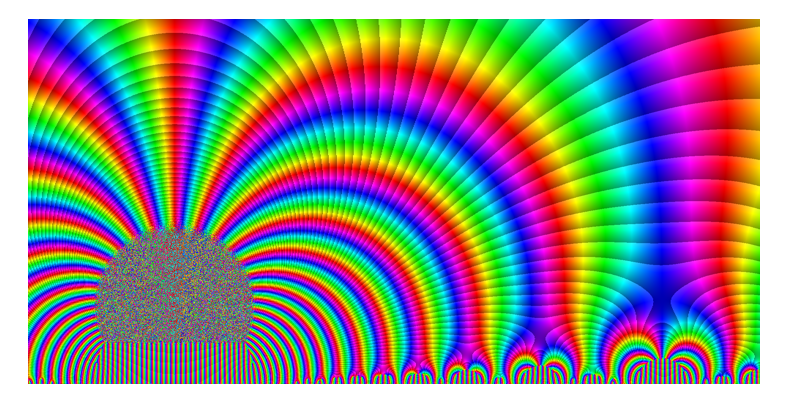}}
\includegraphics[width=0.32\textwidth]{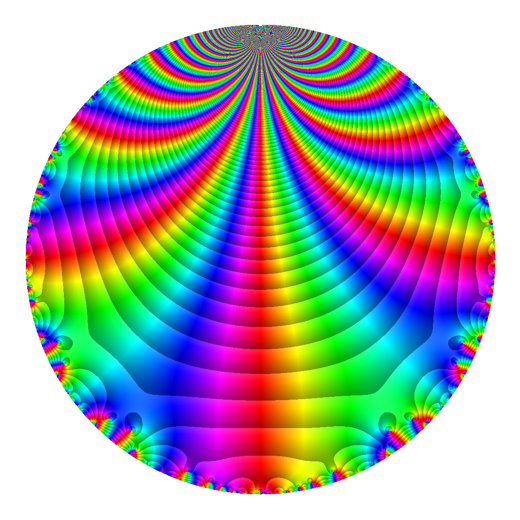}
\raisebox{0.45in}{\includegraphics[width=0.32\textwidth]{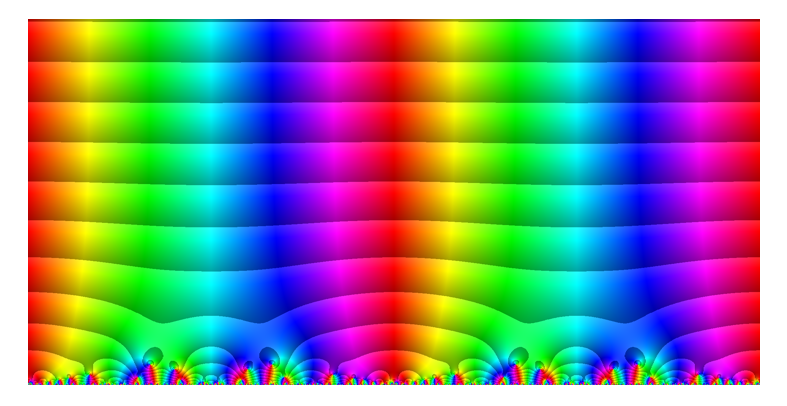}}
\raisebox{0.45in}{\includegraphics[width=0.32\textwidth]{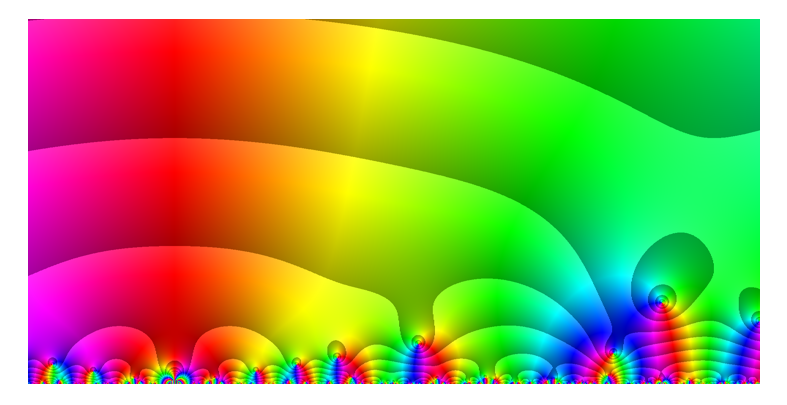}}
\includegraphics[width=0.32\textwidth]{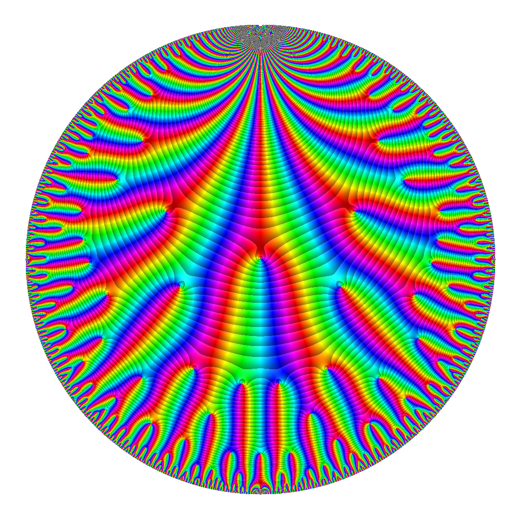}
\includegraphics[width=0.32\textwidth]{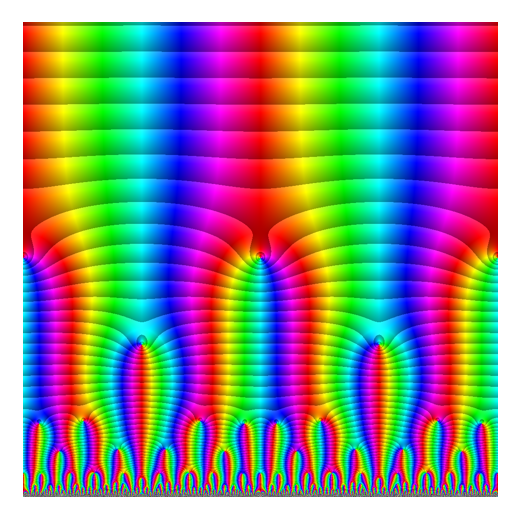}
\raisebox{0.5in}{\includegraphics[width=0.32\textwidth]{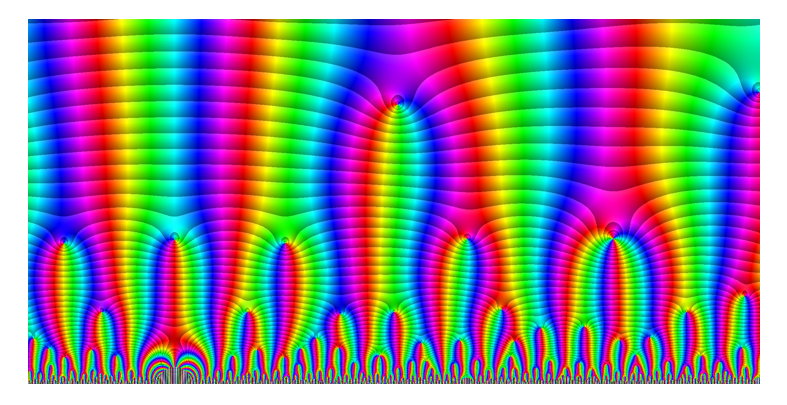}}
\caption{Phase plots with magnitude indicated along contours. Away from the
contours, magnitude is ignored. From top to bottom, the forms are $g, \Delta,
f_{105}$, and $f_{10}$.}\label{fig:phase_contour}
\end{figure}

Again following Wegert~\cite{wegert2010phase}, we create phase plots
with contours representing magnitude. We indicate the contours by adjusting the
lightness: below each contour we adjust the lightness to be darker and above
each contour we adjust the lightness to be lighter. The sudden change from
darker to lighter indicates the contour itself.

The phase naturally maps to hue as in the pure phase plots in
Figure~\ref{fig:phase}. Contours could be chosen with spacing defined linearly
as in~\S\ref{sec:mag_periodic_lin} or logarithmically as
in~\S\ref{sec:mag_periodic_log}. In these plots, we have chosen logarithmic
spacing with respect to $\log_2$. That is, two consecutive contour lines
indicate that the points along one have twice the magnitude of the points along
the other.

\begin{remark}

We emphasize that it is not necessary to compute the contours themselves.
Instead, one defines a coloring of the complex plane that includes brightness
adjustments along the desired contours. In this case, we defined brightness
adjustments around points of magnitude $2^n$ for integers $n$. Then simply
plotting the function with respect to this domain coloring causes the contours
to appear. This is a key idea why approaches based on domain coloring are simple
to compute.

The plotting program used to produce these plots is available
at~\cite{david_lowry_duda_2020_4035117}.

\end{remark}

These plots carry all the information visible in the pure phase plots
of~\S\ref{sec:pure_phase} and some of the information contained within the
logarithmically spaced magnitude plots of~\S\ref{sec:mag_periodic_log}, but
these two sets of information are not displayed equally. The colors grab
attention far more than the brightness adjustments around contours --- analogous
to the lightness difficulties for the standard domain coloring plots
of~\S\ref{sec:standard_domain_coloring}. These plots emphasize argument more
than magnitude.

\section{Choosing Color}\label{sec:color}

We surveyed many different visualizations in~\S\ref{sec:survey}, but one thing
remained constant: the choices of color. In most complex function visualization
packages, there is only one color choice available. All visualizations
in~\S\ref{sec:survey} were created using the only color scheme for
\texttt{complex\_plot} available in SageMath.\footnote{The behavior in Matlab
and Maple appears similar, though it is possible to define a color scheme. On the
other hand Mathematica has an extensive library of complex plotting
color schemes.}

Unfortunately, this color scheme has flaws. For one example, it has uneven
perceptual contrast: it has points of locally high color contrast
(giving the impression of additional local variation) and points of locally low
color contrast (hiding local variation). Thus in Figure~\ref{fig:phase}, for
instance, green is particularly visible and the yellow-green border is
particularly striking, even though there is no underlying feature that this is
identifying. Further, in the plots in the upper halfplane, the vertical bands
of color seem to favor green-yellow at the expense of purple-blue.

In his excellent article~\cite{kovesi2015good}, Kovesi describes this problem
and details how to create colormaps that are perceptually uniform --- that don't
have uneven perceptual contrast. His work led to the creation of new default
colormap, called \emph{viridis}, for Python's \texttt{matplotlib} plotting
library.

More recently, Nu\~nez, Anderton, and Renslow wrote an
article~\cite{nunez2018optimizing} describing their Python module
\texttt{cmaputil}, which helps make perceptually uniform colormaps that
additionally support perception by those with various forms of color vision
deficiency. They created the Python colormap \emph{cividis} (obtained by
modifying \emph{viridis}, hence the name), which they show is a good colormap
for those with reduced sensitivity to red or green light. It is possible to
make other colormaps suitable for other types of perception using this tool.

We have incorporated \texttt{matplotlib} colormaps into our plotting routines.
In particular, we implemented phase plots and phase plots with contours using
these colormaps. In this section, we describe this implementation and give
examples of the resulting plots.

\begin{remark}

SageMath and \texttt{matplotlib} share many plotting routines,
and newer versions of SageMath will default to these new perceptually uniform
colormaps for real-valued plotting. But SageMath's \texttt{complex\_plot} does
not directly use \texttt{matplotlib}.
We are working towards making these new plotting routines easy to use and publicly
available, and possibly incorporating them into SageMath.

\end{remark}

\subsection{Implementation Overview}\label{sec:implementation}

Some of the details of this implementation are specific to Python and SageMath,
but the overall idea readily generalizes.

\begin{enumerate}
\item Begin with a desired complex function $f$, a region $\Omega \subset
\mathbb{C}$, and a desired colormap $\mathrm{cm}$.

\item Evaluate $\arg(f(z))$ and $\lvert f(z) \rvert$ for each point in an $M
\times N$ grid containing $\Omega$. (Optionally mask points $(m, n)$ in this
grid outside of $\Omega$ to avoid unnecessary computation.
For example, to produce the plots of $\Delta$ and $g$ in the disk, we mask
points $z$ with $\lvert z \rvert \geq 1$.)
This gives an $M \times N \times 2$ array.

\item Map each phase to $[0, 1]$. This is most often done through the natural
map $\theta \mapsto (\theta/2\pi) \bmod 1$, but note that it is possible to
add an offset $\Theta$ through $\theta \mapsto (\theta/2\pi + \Theta) \bmod 1$.
This might be done to choose which phase corresponds to the endpoints of the
colormap. (See plots in Figure~\ref{fig:color_general} for examples).

\item Each \texttt{matplotlib} colormap is implemented as a function
$\mathrm{cm}: [0, 1] \rightarrow \mathrm{RGBA}$, where $\mathrm{RGBA} \cong
{[0, 1]}^4$ is red-green-blue-alpha colorspace. We don't currently use the
alpha channel. Thus apply $\mathrm{cm}$ to each mapped phase and discard the
alpha channel to get $\mathrm{RGB}$.

\end{enumerate}

If the intention is to produce a pure phase plot, then (after discarding the
magnitude) one now has an $M \times N \times \mathrm{RGB}$ array, and one now
produces a plot using this data in this grid, either directly or using a
plotting library. For example, \texttt{matplotlib}'s method
\texttt{matplotlib.pyplot.imshow} will directly plot this array.

However, if the intention is to produce a phase plot with (logarithmically
spaced) magnitude contours, then additional work is necessary. We would like to
adjust the brightness as done in~\S\ref{sec:phase_mag}, but the brightness of an
$\mathrm{RGB}$ pixel is nontrivial to work with directly.

Instead we convert each pixel to $\mathrm{HSL}$ colorspace
(hue-saturation-lightness) and adjust lightness. As most plotting libraries use
$\mathrm{RGB}$ (as does the underlying hardware), we convert back to
$\mathrm{RGB}$ prior to plotting. Thus to produce a phase plot with magnitude
contours we perform these additional steps.

\begin{enumerate}
\setcounter{enumi}{4}
\item Map each $\mathrm{RGB}$ element to $\mathrm{HSL}$ colorspace, where
$\mathrm{HSL}$ refers to ``hue-saturation-lightness''.
In $\mathrm{HSL}$, it is possible to directly manipulate the lightness. This
gives an $M \times N \times 3$ array.

\item Map each magnitude to a lightness adjustment in the range $[-1, 1]$. For
example, to have logarithmic contours, one could use a map of the form
$(\log_2(\lvert f(z) \rvert) \bmod 1)/2$ with a fixed lightness adjustment in a
neighborhood of $0$. This has the effect of creating logarithmically spaced
contours. This gives an $M \times N \times 1$ array.

We note that we refer to $x \bmod 1$ to mean the least
nonnegative number of the form $x + \ell$ for $\ell \in \mathbb{Z}$, i.e.\ the
``computer science'' modulo operation.

\item For each $(m, n)$, apply the corresponding lightness adjustment to each
$\mathrm{HSL}$ value. (It may be necessary to renormalize lightness at this
step). This gives a single $M \times N \times 3$ array.

\item Map each $\mathrm{HSL}$ back to $\mathrm{RGB}$ to get the final $M \times
N \times 3$ array.
\end{enumerate}

And now one can directly plot this array or use a library like
\texttt{matplotlib} to plot this array, as noted above.

\begin{remark}

Each step can be done very quickly if one uses vectorized array operations as in
NumPy~\cite{numpy}.
We also note that Python's \texttt{colorsys} library implements NumPy-vectorized
conversions between $\mathrm{RGB}$ and $\mathrm{HSL}$.
Then the most time-consuming step is then the evaluation of $f$
on the $M \times N$ grid, as should be expected.

\end{remark}

\subsection{Example visualizations with colormaps}

We now give example visualizations with \texttt{matplotlib} colormaps. The
three mostly commonly used \texttt{matplotlib} colormaps are currently
\emph{twilight}, \emph{viridis}, and \emph{cividis}. Each of these maps is
perceptually uniform and designed for clean presentation of data.

\begin{figure}[!ht]
\centering
\includegraphics[width=0.32\textwidth]{g2_phase.png}
\includegraphics[width=0.32\textwidth]{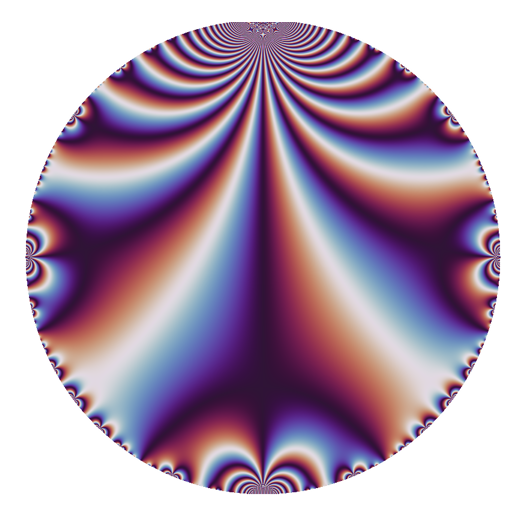}

\includegraphics[width=0.32\textwidth]{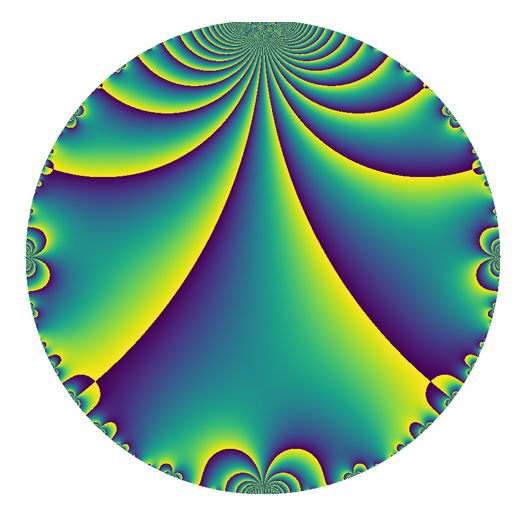}
\includegraphics[width=0.32\textwidth]{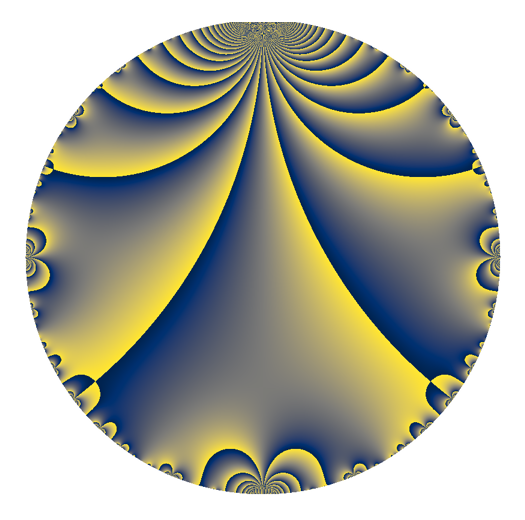}
\caption{
Phase plots of $g$ with four color schemes. The top left is the
default color scheme of SageMath. The top right is \emph{twilight}, bottom left is
\emph{viridis}, and the bottom right is \emph{cividis}.}\label{fig:color_compare}
\end{figure}
%

The \emph{twilight} colormap is also cyclic. This means that the two ends of
the colormap line up perfectly, and thus it is a good colormap to represent
phase in phase plots. 
The \emph{viridis} and \emph{cividis} colormaps are not cyclic. There is a
sharp difference in colors at the two ends of the colormap . Plotting a modular
form using either \emph{viridis} or \emph{cividis} thus highlights the points
where the phase crosses over the end of the colormap.

In some cases, this may be undesirable. For exploratory visualization, using a
cyclic perceptually uniform colormap often seems superior. But more generally
it is sometimes desirable, or even advantageous, to emphasize certain aspects
of a complex plot, and using a non-cyclic colormap to highlight phase-changes
is just one possible method.

\begin{figure}[!ht]
\centering
\includegraphics[width=0.32\textwidth]{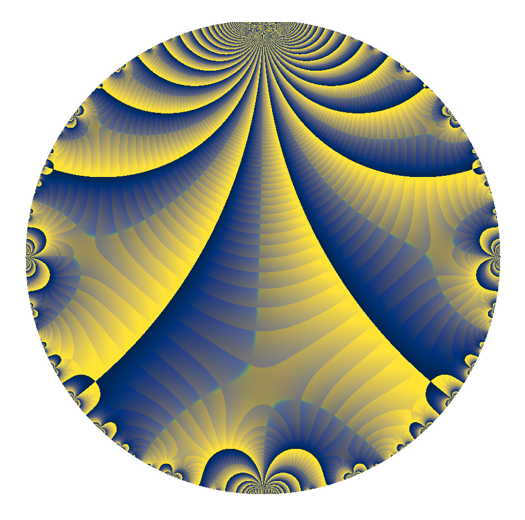}
\raisebox{0.5in}{\includegraphics[width=0.32\textwidth]{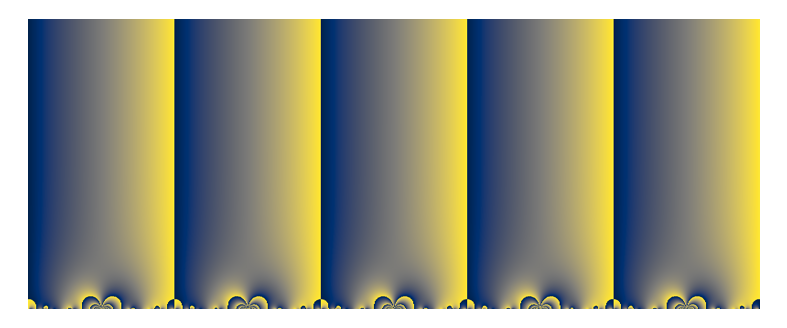}}
\raisebox{0.5in}{\includegraphics[width=0.32\textwidth]{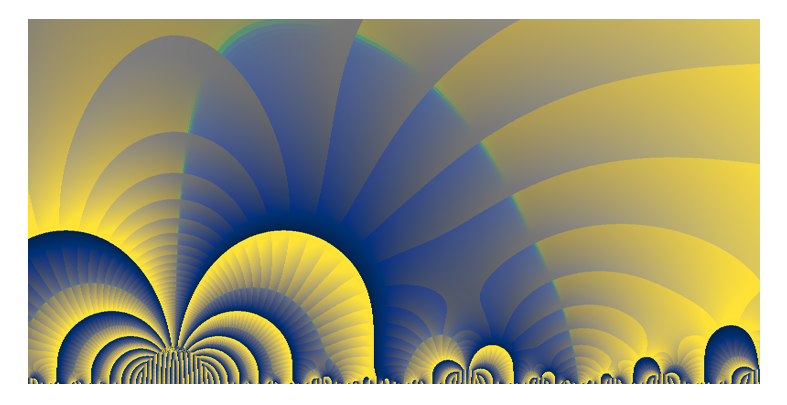}}
\includegraphics[width=0.32\textwidth]{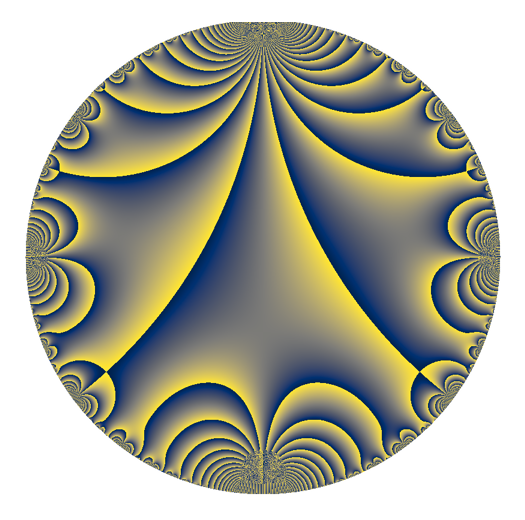}
\includegraphics[width=0.32\textwidth]{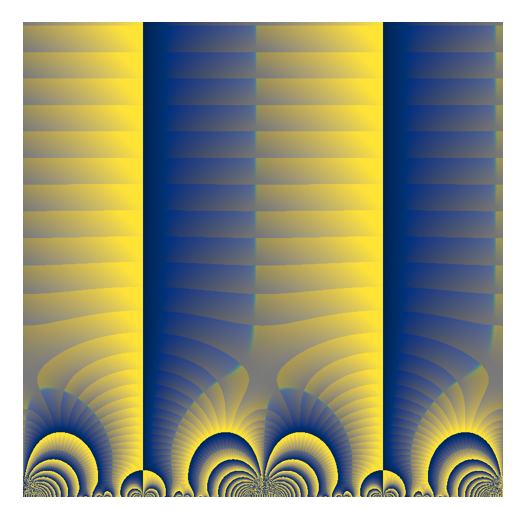}
\raisebox{0.5in}{\includegraphics[width=0.32\textwidth]{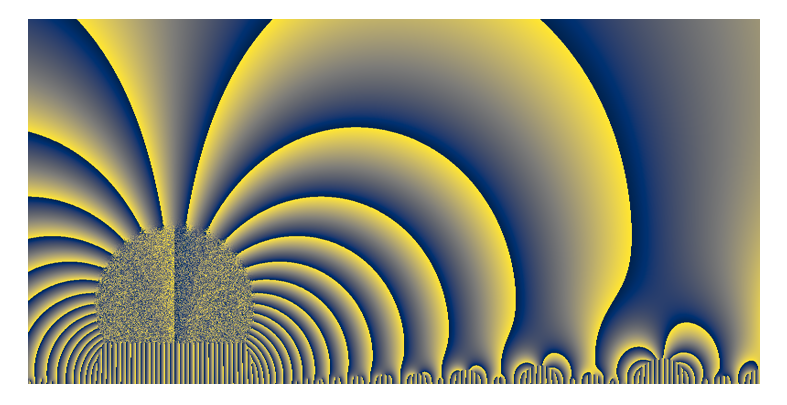}}
\includegraphics[width=0.32\textwidth]{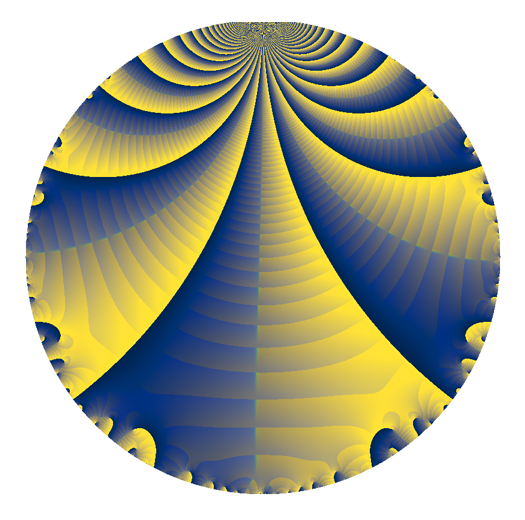}
\raisebox{0.45in}{\includegraphics[width=0.32\textwidth]{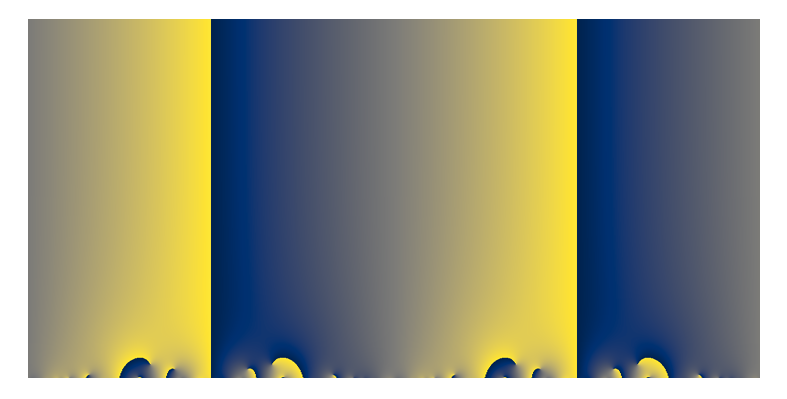}}
\raisebox{0.45in}{\includegraphics[width=0.32\textwidth]{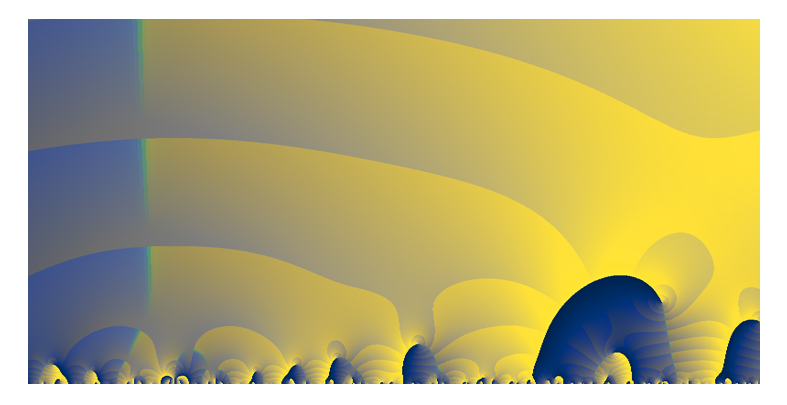}}
\includegraphics[width=0.32\textwidth]{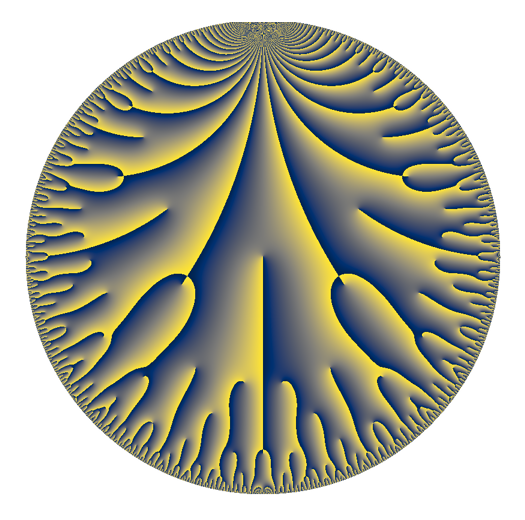}
\includegraphics[width=0.32\textwidth]{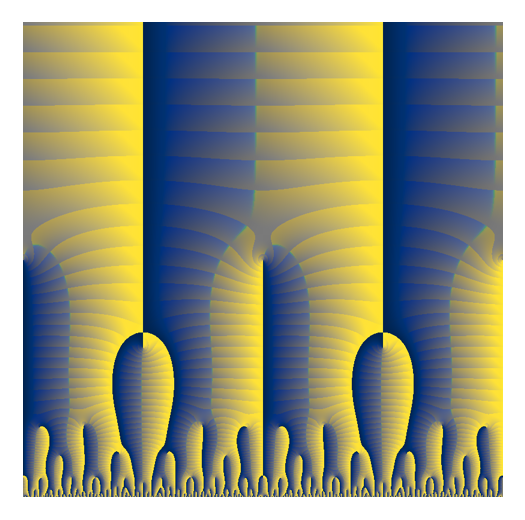}
\raisebox{0.5in}{\includegraphics[width=0.32\textwidth]{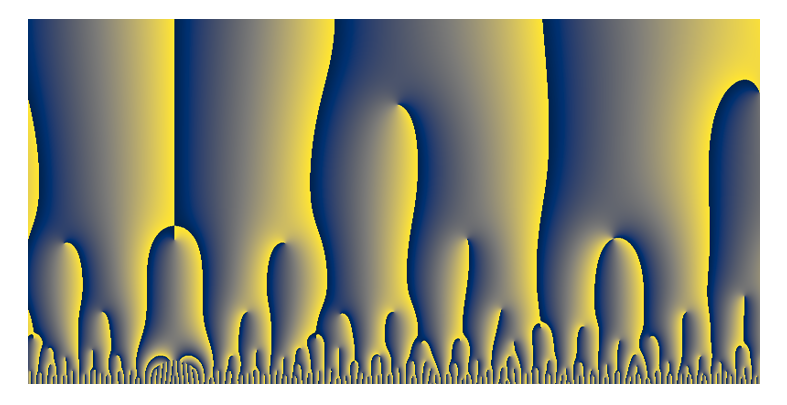}}
\caption{
Phase plots with the \texttt{matplotlib} perceptually uniform colormap
\emph{cividis} applied. In half of the plots, logarithmically spaced contours
are included. From top to bottom, the forms are $g, \Delta,
f_{105}$, and $f_{10}$.}\label{fig:cividis}
\end{figure}

In Figure~\ref{fig:color_compare}, we plot the modular form $g$ in four
different color schemes: the default in SageMath, and the three
\texttt{matplotlib} colormaps \emph{twilight}, \emph{viridis}, and
\emph{cividis}. In contrast to the default color scheme, these three colormaps
produce a plot that is more balanced in appearance. In particular, the
perceptually uniform plots don't overemphasize regions like the bright green
and yellow portions in the default color scheme.

Of course, perceptually uniform doesn't mean perfect. In the \emph{twilight}
colormap, there is a sensation of peaks and troughs resulting from the light
and dark areas of the plots; and as both \emph{viridis} and \emph{cividis}
aren't cyclic, the sharp color change at the edges of the colormap gives the
impression of contours. These are features of the colormap, not of the
underlying function.

In Figure~\ref{fig:cividis}, we give phase plots in the \emph{cividis}
colormap. For half of the plots, we include logarithmically spaced contours,
implemented as described above. These plots should compared with the phase
plots in Figure~\ref{fig:phase} and the contoured phase plots in
Figure~\ref{fig:phase_contour}.

In these images, we have mapped the arguments to the colormap in such a way
that the sharp change in color from light yellow to dark blue occurs when the
phase is $\pi$, or equivalently when the value of the modular form is a
negative real number. On the upper halfplane, a modular form with real
coefficients will take real values on vertical lines $z = \frac{n}{2} + it$ for
$n \in \mathbb{Z}, t > 0$. This explains the regularly spaced lines in the
plots.

On the one hand, this gives a strong and simple visual indicator of phase
change. The regularity of this phase change is more obvious in these plots than
the previous phase plots in the default color scheme. But on the other hand,
this strongly emphasizes a small subset of phases.

We note also that the colormap \emph{cividis} is both visually appealing and
widely accessible.

\begin{figure}[!t]
\centering
\includegraphics[width=0.32\textwidth]{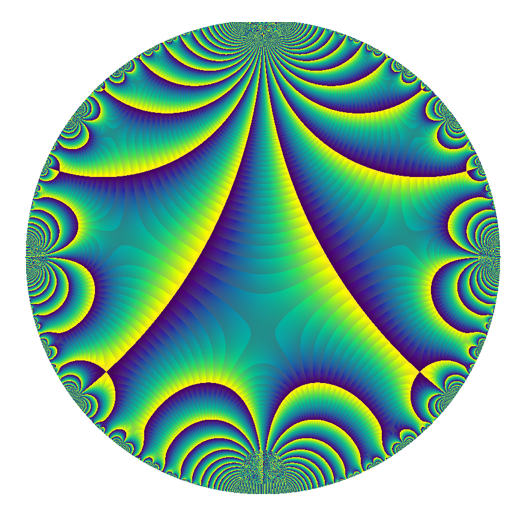}
\includegraphics[width=0.32\textwidth]{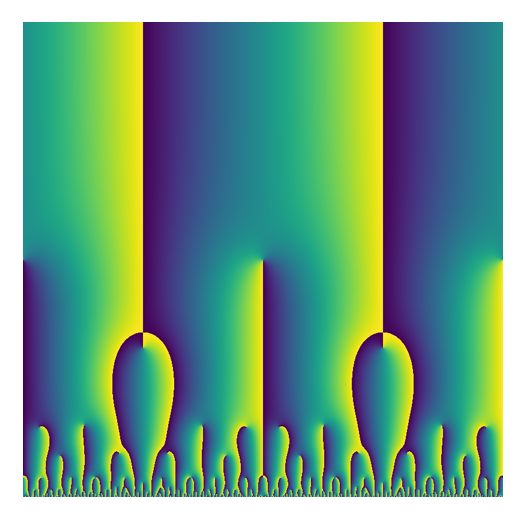}
\raisebox{0.5in}{\includegraphics[width=0.32\textwidth]{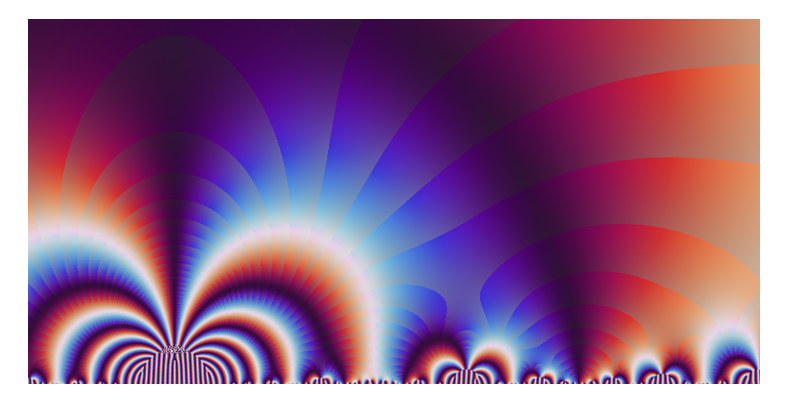}}

\includegraphics[width=0.32\textwidth]{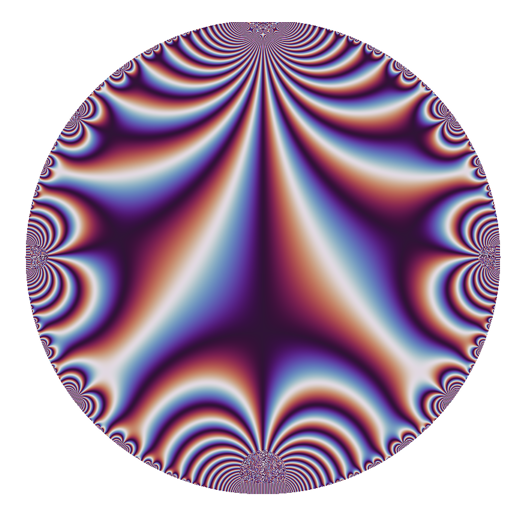}
\includegraphics[width=0.32\textwidth]{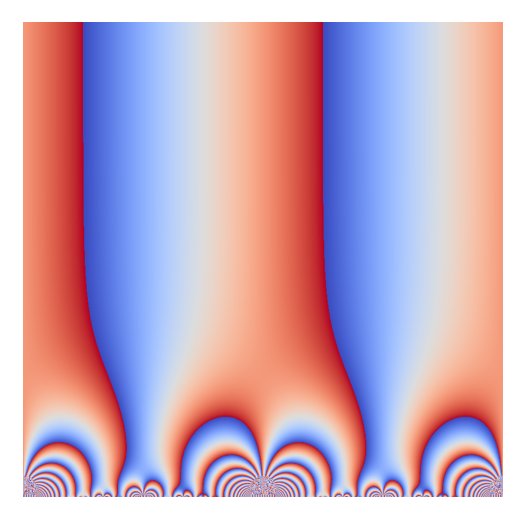}
\raisebox{0.5in}{\includegraphics[width=0.32\textwidth]{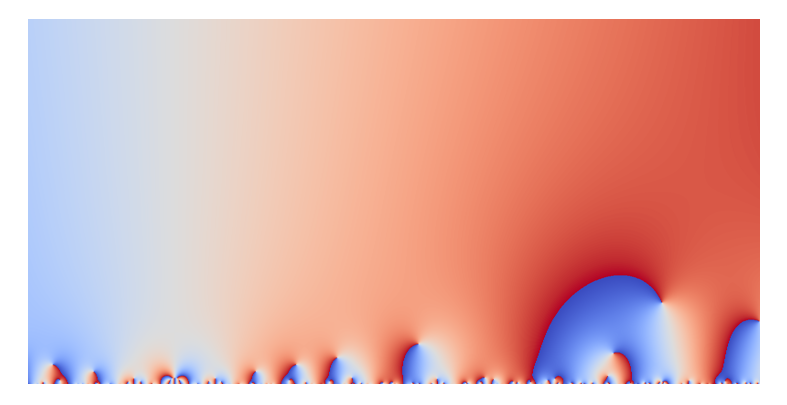}}

\includegraphics[width=0.32\textwidth]{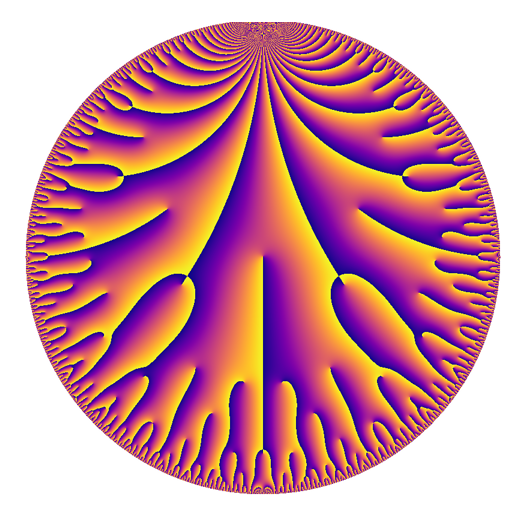}
\raisebox{0.5in}{\includegraphics[width=0.32\textwidth]{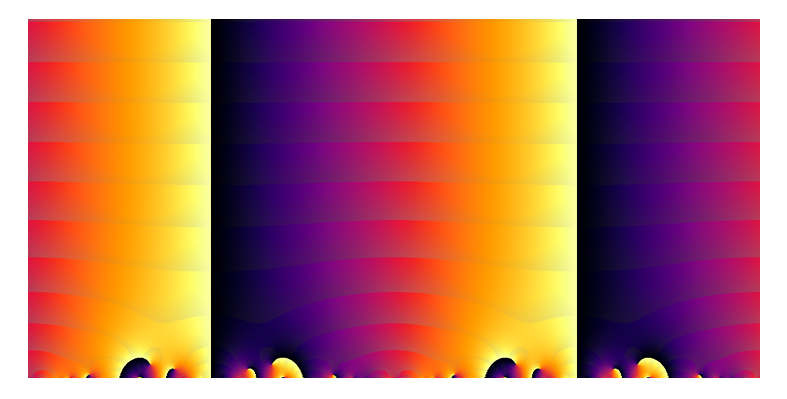}}
\raisebox{0.5in}{\includegraphics[width=0.32\textwidth]{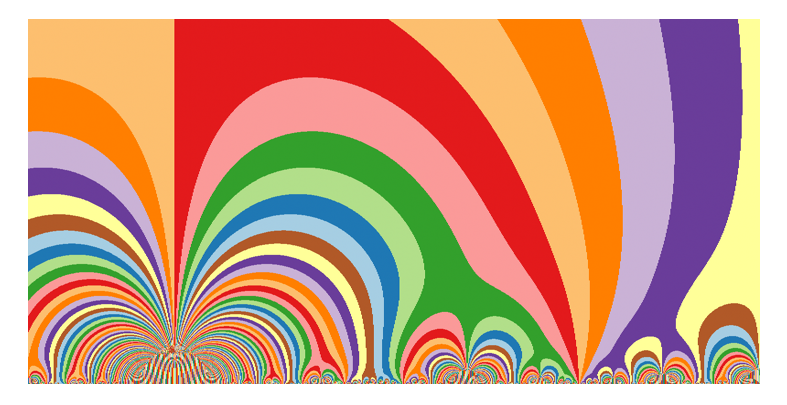}}
\caption{%
Phase plots with a variety of colormaps from \texttt{matplotlib} applied.
From the top left and going row by row, the (function, colormap) are
($\Delta$, \emph{viridis}), ($f_{10}$, \emph{viridis}), ($g$, \emph{twilight});
($\Delta$, \emph{twilight}), ($\Delta$, an offset \emph{coolwarm}), ($f_{105}$, \emph{coolwarm});
($f_{10}$, \emph{plasma}), ($f_{105}$, \emph{inferno}), ($g$, \emph{Paired}).
}\label{fig:color_general}

\includegraphics[width=0.32\textwidth]{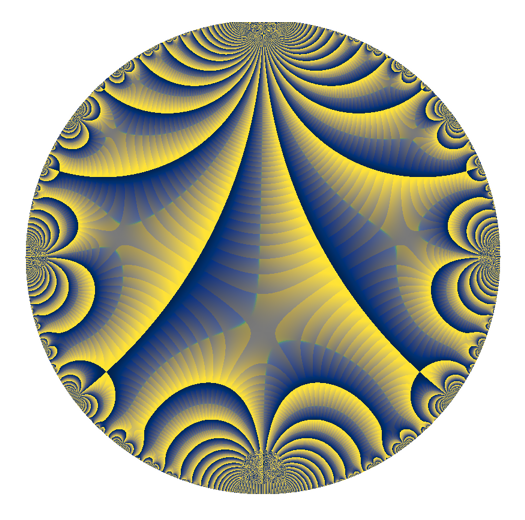}
\includegraphics[width=0.32\textwidth]{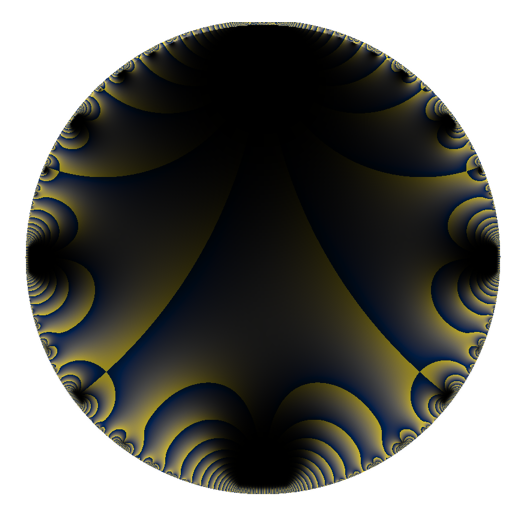}
\caption{%
Two visualizations of $\Delta$.}\label{fig:example_pair}
\end{figure}

In Figure~\ref{fig:color_general} we have included plots in a variety of other
\texttt{matplotlib} colormaps. For the top six plots, we've used the three
colormaps \emph{viridis}, \emph{twilight}, and \emph{coolwarm}.
Note that in the middle plot in the second row, we have normalized the
non-cyclic colormap \emph{coolwarm} so that the sharp distinction between red
and blue occurs when the argument of $\Delta$ is $3 \pi / 2$ instead of $\pi$.
In the bottom row, we've applied the colormaps \emph{inferno}, \emph{plasma},
and \emph{Paired}. Both \emph{inferno} and \emph{plasma} are also perceptually
uniform, but are very orange. The colormap \emph{Paired} is a discrete
colormap, which has the effect of coloring different regions of the complex
plane based on discrete argument sectors.

While each plot in Figure~\ref{fig:color_general} may not be appropriate for
initial exploratory visualization, they demonstrate the ability to choose
colormap to highlight certain behaviors or areas of the function. And most
importantly, they demonstrate that implementing \texttt{matplotlib} colormaps
gives access to an existing, very large library of color schemes.

\begin{remark}

Choosing color schemes to highlight certain behaviors is not a new concept.
For example, Frank Farris has produced several pieces of art based on domain
colorings. See his website~\url{http://math.scu.edu/~ffarris/} for more.

\end{remark}

Finally, we note that it is possible to apply these colormaps to the previous
styles of visualization. For example, we can produce a colormapped version of
the default SageMath plot (as in Figure~\ref{fig:SDC_1}) by changing steps
(5)--(8) in the implementation described in~\ref{sec:implementation}; map the
phase to RGB as described there, but map magnitude to brightness as in
\S\ref{sec:standard_domain_coloring} or \S\ref{sec:mag_no_color}.

\subsection*{A strategy for making visualizations}

Having given examples of several different types of complex plot, we feel we
should describe our current preferences for making plots. If we were to choose
a single plotting method to present a visualization, we would use contoured
phase plots as in Figure~\ref{fig:cividis}.

We have found that for initial exploration, it is very
effective to pair a contoured phase plot as in Figure~\ref{fig:cividis} with a
plot showing the absolute magnitude, such as a colormapped version of the
default Sagemath plot.

In Figure~\ref{fig:example_pair}, we do this for $\Delta$ on the disk in the
\emph{cividis} colormap. The strengths and weaknesses of these two plots
complement each other. Specifically, the relative magnitudes are very clearly
given in the (colormapped) standard domain coloring plot on the right and the
phase information is more apparent in the contoured phase plot on the left. The
presence of phase information in the standard domain coloring helps to identify
the same location in the two plots.

After this initial investigation, it is possible to
choose regions and colormaps to highlight different behaviors of the function.

But overall, the idea of choosing colormaps for exploratory complex function
visual analysis is young, and there remains much to be discovered.

%
%

\bibliographystyle{alpha}
\bibliography{bibfile}

\begin{thebibliography}{{LMF}19}

\bibitem[DS05]{diamond2005first}
Fred Diamond and Jerry Shurman.
\newblock {\em A first course in modular forms}, volume 228.
\newblock Springer Verlag, 2005.

\bibitem[Far98]{farris1998visual}
Frank~A Farris.
\newblock Visual complex analysis. by tristan needham.
\newblock {\em The American Mathematical Monthly}, 105(6):570--576, 1998.

\bibitem[Hun07]{matplotlib}
J.~D. Hunter.
\newblock Matplotlib: A 2d graphics environment.
\newblock {\em Computing in Science \& Engineering}, 9(3):90--95, 2007.

\bibitem[Kov15]{kovesi2015good}
Peter Kovesi.
\newblock Good colour maps: How to design them.
\newblock {\em arXiv preprint arXiv:1509.03700}, 2015.

\bibitem[LD20]{david_lowry_duda_2020_4035117}
David Lowry-Duda.
\newblock phase\_mag\_plot.
\newblock \url{https://github.com/davidlowryduda/phase_mag_plot/}, September
  2020.
\newblock [Online; Reference version at
  \url{https://doi.org/10.5281/zenodo.4035117}].

\bibitem[{LMF}19]{lmfdb}
The {LMFDB Collaboration}.
\newblock The {L}-functions and modular forms database.
\newblock \url{http://www.lmfdb.org}, 2019.
\newblock [Online; accessed 30 October 2019].

\bibitem[NAR18]{nunez2018optimizing}
Jamie~R Nu{\~n}ez, Christopher~R Anderton, and Ryan~S Renslow.
\newblock Optimizing colormaps with consideration for color vision deficiency
  to enable accurate interpretation of scientific data.
\newblock {\em PloS one}, 13(7):e0199239, 2018.

\bibitem[Oli06]{numpy}
Travis~E Oliphant.
\newblock {\em A guide to NumPy}, volume~1.
\newblock Trelgol Publishing USA, 2006.

\bibitem[{Sag}20]{sage}
{Sage Developers}.
\newblock {\em {S}ageMath, the {S}age {M}athematics {S}oftware {S}ystem
  ({V}ersion 8.8)}, 2020.
\newblock {\tt https://www.sagemath.org}.

\bibitem[WS10]{wegert2010phase}
Elias Wegert and Gunter Semmler.
\newblock Phase plots of complex functions: a journey in illustration.
\newblock {\em Notices AMS}, 58:768--780, 2010.

\end{thebibliography}

\end{document}